\documentclass[aps,showpacs]{revtex4}
\usepackage{graphicx}
\usepackage{float}
\usepackage[T1]{fontenc}
\usepackage[latin1]{inputenc}
\usepackage{amssymb}
\include{epsf}
\epsfverbosetrue

\newcommand{\xx}{\noindent}
\newcommand{\ra}{\rightarrow}
\newcommand{\Sign}{{\rm Sign}}

\newcommand{\Psl}{P\!\!\!\!/~}
\newcommand{\Ksl}{K\!\!\!\!/~}

\newcommand{\bea}{\begin{eqnarray}}
\newcommand{\eea}{\end{eqnarray}}

\makeatother
\begin{document}
\setlength{\unitlength}{1mm}

\title{\Large{Leading Order QCD Shear Viscosity from the 3PI Effective Action}}

\author{M.E. Carrington and E. Kovalchuk}

\email{carrington@brandonu.ca; kavalchuke@brandonu.ca}
 \affiliation{Department of Physics, Brandon University, Brandon, Manitoba, R7A 6A9 Canada\\ and \\
  Winnipeg Institute for Theoretical Physics, Winnipeg, Manitoba }

\begin{abstract}
In this article we calculate the leading order shear viscosity in QCD using the resummed 3PI effective action. 
We work to 3-loop order in the effective action. We show that the integral equations that resum the pinching and collinear contributions are produced naturally by the formalism. All leading order terms are included,  without the need for any kind of power counting arguments.
\end{abstract}

\pacs{11.15.-q, 11.10.Wx, 05.70.Ln, 52.25.Fi}
\maketitle
\section{Introduction}

Transport coefficients measure the efficiency with which conserved quantities are transported through a medium, over distances that are long compared to the microscopic relaxation scales of the system (for a review see \cite{GA-review}). Direct applications include the early universe and the quark gluon plasma. The calculation of transport coefficients is also important from a purely theoretical standpoint. They characterize linear deviations from equilibrium, but are calculated using the familiar methods of equilibrium field theory. Results could therefore be used as a check on calculations based on purely non-equilibrium methods. 

The calculation of transport coefficients in gauge theories is hugely complicated by the occurrance of pinch- and collinear-singularities \cite{AMY1,AMY2,AMY3}. These singularities can be regulated by using hard thermal loop propagators. However, this remedy produces infinite sets of graphs which all contribute at the same order. In order to include all leading order contributions, these infinite sets of graphs need to be resummed. The problem is to develop a technique to perform the resummation that avoids double counting and respects all of the symmetries of the original theory. 

The complete leading order calculation of electrical conductivity and shear viscosity, in both QED and QCD, was done in \cite{AMY3}. This calculation is not obtained directly from quantum field theory, but is derived from kinetic theory. It is of interest to understand the connection between the kinetic theory approach, and a calculation based on quantum field theory. One motivation is that quantum field theory might provide a better framework than kinetic theory for calculations beyond leading order.  The equivalence of the two approaches has been demonstrated for scalar theories \cite{jeonPhi4,jeonYaffe,MC-houRK,enkeHeinz}. In QED, field theory based calculations have been done using a direct ladder summation \cite{basa,gertWI}, dynamical remormalization group methods \cite{rg}, and other diagrammatic methods \cite{jeonSigma,jeonEta}. A large $N_f$ leading log  calculation of conductivity and shear viscosity was done in \cite{gertNf} using the 2PI effective theory. The conductivity has been obtained at leading order from the 3PI effective action \cite{MC-EK1,MC-EK2}. 
In QCD, the field theoretic calculation of shear viscosity has only been done at leading log order \cite{basa,houQCD}. In this paper we present the first  calculation of the full leading order QCD shear viscosity using quantum field theory methods. We show that the calculation can be organized naturally using  the 3PI effective action. The result provides strong support for the use of $n$PI effective theories as a method to study the equilibration of quantum fields. 

The $n$-PI effective theory, in which the $n$-point functions are treated as variational parameters, is a natural method to organize the calculation of transport coefficients. In general, a consistent resummation requires the computation of the $n$PI effective action for infinite $n$. However, there is an equivalence heirarchy that simplifies the structure of the calculation \cite{berges1}: to 2-loop order, the infinite-PI effective action is equivalent to the 2PI effective action, to 3-loop order, the infinite-PI effective action is equivalent to the 3PI effective action, etc. 

Truncations of the effective action produce problems with gauge invariance. Even though the effective action is consistent with the global symmetries of the theory, the Ward identities associated with the gauge symmetry may not be satisfied for the self consistently determined vertex functions \cite{smit,HZ}. To address this problem, we use the resummed effective action, which is defined with respect to the self-consistent solutions of the $n$-point functions \cite{baym,vanh, reinosa1,reinosa2}. 

There are three different types of $n$-point functions involved in the calculation: (1) the self-consistent solutions of the equations of motion; (2) `mixed' $n$-point functions obtained by taking functional derivatives of the self-consistent solutions with respect to field expectation values and; (3) `external' $n$-point functions obtained by taking functional derivatives of the resummed effective action with respect to field expectation values. For QED, it has been shown that the `external' $n$-point functions satisfy the usual Ward identities \cite{MC-EK1,serreau}. At the exact level, all of these definitions are equivalent to each other. Integral equations for type (1) vertices are obtained from the equations of motion of the effective action (see Eqn. (\ref{eom1})). Integral equations for type (2) vertex functions are obtained by functionally differentiating the equations of motion with respect to field expectation values (see, for example, Eqn. (\ref{omegaSt})).

This paper is organized as follows. 
In section \ref{Notation} we define some notation. 
In section \ref{eta} we derive an expression for the shear viscosity in terms of the integrand that gives the gluon polarization tensor.
In section \ref{section3PI} we give the 3PI effective action to 3-loop order.
In section \ref{sectionEXT} we define some `external' $n$-point functions and obtain an expression for the `external' 2-point function. 
In sections \ref{sectionBS} and \ref{sectionEOM}  we derive integral equations for the relevant `mixed' vertices, and the self consistent vertices. 
In section \ref{sectionME} we show that the kernels of these equations can be written as the square of the sum of the amplitudes that correspond to all physical scattering and production processes. 
In section \ref{sectionCONC} we present our conclusions.
In Appendix A we define the notation used in the Keldysh representation of finite temperature field theory.
In Appendix B we show that the expansion of the `external' polarization tensor produced by the 3PI formalism contains all of the terms that would be produced by a Wick expansion.  In Appendix C we give some details of the calculation presented in section \ref{sectionME}.

\section{Notation}
\label{Notation}

We use:
\bea
\label{not1}
&& g={\rm diag}(1,-1,-1,-1)\,, \\
&& n_b(p_0) = \frac{1}{e^{\beta p_0}-1}\,,~~n_f(p_0) = \frac{1}{e^{\beta p_0}+1}\,, \nonumber\\
&&N_B(p_0) = 1+2n_b(p_0)\,,~~N_F(p_0) = 1-2n_f(p_0)\,, \nonumber\\
&&\int dP:= \int \frac{d^4p}{(2\pi)^4}\,,~~\int_p:= \int \frac{d^3p}{2E_p\,(2\pi)^3}\,,\nonumber\\
&&\hat I^{ij}=\big(p^i p^j-\frac{1}{3}p^2\delta^{ij}\big)\,\frac{1}{p^2}\,.\nonumber
\eea
We work in the Feynman gauge and write the QCD Lagrangian as:
\bea
\label{lagrangian}
&&{\cal L}=-\frac{1}{4}F^a_{\mu\nu}F^{\mu\nu ~ a}-\frac{1}{2}(\partial^\mu A_\mu^a)^2+ i \bar \psi \gamma_\mu D^\mu  \psi -\bar\eta^a \partial^\mu (D_\mu \eta)^a \,,\\[2mm]
&& F_{\mu\nu}^a=\partial_\mu A_\nu^a-\partial_\nu A_\mu^a+g f^{abc}A^b_\mu A^c_\nu\,, \nonumber\\
&& D_\mu  \psi = (\partial_\mu-i g A^a_\mu t^a)\psi\,,\nonumber\\
&& (D_\mu  \eta)^a = \big(\partial_\mu \delta^{ac}+g f^{abc}A_\mu^b\big)\eta^c\,.\nonumber
\eea
The classical action is:
\bea
S_{cl}[\psi,\bar\psi,A,\eta,\bar\eta]=\int d^4x \,{\cal L}\,.
\eea
The group factor notation for $SU(N)$ is:
\bea
{\rm fundamental~representation}:~~&&C_F=(N^2-1)/(2N)\,,~T_F=1/2\,,~d_F=N\,,\\[1mm]
{\rm adjoint~representation}:~~&& C_A=N\,,~T_A=N\,,~d_A=N^2-1\,.\nonumber
\eea
For simplicity we will set the coupling constant $g$ to one throughout.

\section{Shear Viscosity}
\label{eta}

The Kubo formula for shear viscosity is:
\bea
\label{visco}
&&\eta=\frac{1}{20}\left(\frac{\partial}{\partial q_0}2\, {\rm Im}\,\rho_{\pi\pi}(q_0,0)\right)\Big|_{q_0\rightarrow 0}\\[2mm]
&&\rho_{\pi\pi}(q_0,0)=\int^\infty_{-\infty}dt\,\int d^3x\, e^{iq_0t}\,\theta(t)\,\langle \pi^{ij}(t,x),\pi^{ij}(0)\rangle\nonumber
\eea
where $\pi^{ij}$ is the traceless part of the energy momentum tensor.

We begin by considering the contribution to the shear viscosity that corresponds to using bare propagators and vertices. We will call this quantity $\bar\eta$. We use the Keldysh representation of the real time formulation of finite temperature field theory. The basic method is described in Appendix A. The derivation of the corresponding expression for the case of the QED conductivity is given in detail in \cite{MC-EK1}. We obtain$^1$\footnotetext[1]{Our notation throughout this paper differs slightly from that used in \cite{MC-EK1,MC-EK2}. In this paper, the symbols used to represent propagators and vertices correspond directly to lines and dots in diagrams, with no additional factors of $\pm i$.}
\bea
\label{viscoInt}
&&\bar\eta = -\frac{1}{10}\beta \int dP
\bigg[n_f(p_0)\big(1-n_f(p_0)\big){\rm Tr}\bigg((\Lambda_0)_{cc^\prime}^{ij}(P+Q,Q,P)S^0(P+Q)_{ret}\,(\Lambda_0)_{cc^\prime}^{ij}(P+Q,Q,P)S^0(P)_{adv}\bigg)\\
&&
+n_b(p_0)\big( 1+n_b(p_0)\big)\bigg(\frac{1}{2}\, (\Omega_0)_{ab}^{ij\lambda\tau}(-P-Q,Q,P)D^0_{\tau\tau^\prime}(P+Q)_{ret}\,(\Omega_0)^{ij\tau^\prime\lambda^\prime}_{ab}(-P-Q,Q,P)\,D^0_{\lambda\lambda^\prime}(P)_{adv}\, \nonumber\\
&&
-\,(\Theta_0)_{ab}^{ij}(P+Q,Q,P)G^0(P+Q)_{ret}\,(\Theta_0)_{ab}^{ij}(P+Q,Q,P)G^0(P)_{adv}\bigg)  \bigg]_{\scriptsize{
\begin{array}{l}
\vec q =0 \\
q_0\to 0
\end{array}
}}\nonumber
\eea
The vertices are defined as:
\bea
\label{vertTrans}
&&(2\pi)^4 \delta^4(K+P+Q)(\Omega_0)^{ij\lambda\tau}_{ab}(K,Q,P)=\int d^4 x\int d^4 y\int d^4 z\,e^{-i q x}e^{-i p y}e^{-ik z}\langle A^\lambda_a(z)\pi^{ij}_{\rm gluon}(x) A^\tau_b(y)\rangle \,,\\
&&(2\pi)^4 \delta^4(K-P-Q)(\Theta_0)_{ab}^{ij}(K,Q,P)=\int d^4 x\int d^4 y\int d^4 z\,e^{-i q x}e^{-i p y}e^{ik z}\langle \eta_a(z)\pi^{ij}_{\rm ghost}(x) \bar\eta_b(y)\rangle \,,\nonumber\\
&&(2\pi)^4 \delta^4(K-P-Q)(\Lambda_0)_{cc^\prime}^{ij}(K,Q,P)=\int d^4 x\int d^4 y\int d^4 z\,e^{-i q x}e^{-i p y}e^{ik z}\langle \psi_c(z)\pi^{ij}_{\rm quark}(x) \bar\psi_{c^\prime}(y)\rangle \,,\nonumber
\nonumber
\eea
where $\{\pi^{ij}_{\rm quark},~\pi^{ij}_{\rm ghost},~\pi^{ij}_{\rm gluon}\}$ indicates the part of the traceless energy momentum tensor that is quadratic in quark, ghost, and gluon fields. For the ghost and quark vertices the order of the momenta is:  outgoing momentum of the outgoing ghost/quark, incoming momentum of the gluon, incoming momentum of the incoming ghost/quark. For the gluon vertex all momenta are incoming. 

\begin{figure}[H]
\begin{center}
\includegraphics[width=2cm]{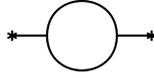}
\end{center}
\caption{A generic contribution to the shear viscosity.}
\label{figNai}
\end{figure}

The integral in (\ref{viscoInt}) contains three bubble-type diagrams: a gluon bubble, a ghost bubble, and a quark bubble. These diagrams contain pinching and collinear singularities that need to be resummed in order to obtain the full leading order contribution. 
In order to simplify the explanation of this point, we consider a generic bubble diagram, and draw the propagators as solid lines, as shown in Fig. \ref{figNai}.  Throughout this paper we will use stars to indicate the legs of a type (2) or type (3) $n$-point function that correspond to functional derivatives with respect to field expectation values (see also Fig. \ref{figExt-V}).
\par

\begin{figure}[H]
\begin{center}
\includegraphics[width=10cm]{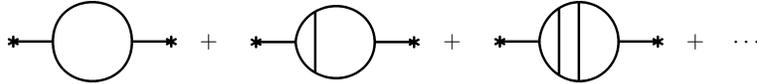}
\end{center}
\caption{Contributions to the shear viscosity from pinching singularities.}
\label{figPinch}
\end{figure}

Pinch singularities are produced by the low frequency limit in the Kubo formula (Eqn. (\ref{visco})).  When integrating a term of the form $\int dp_0 \;G^{ret}(P)G^{adv}(P)$, the integration contour is `pinched' between poles on each side of the real axis, and the integral contains a divergence called a `pinch singularity.' 
A set of graphs containing pinch singularities has the general form shown in Fig. \ref{figPinch}.
\par

\begin{figure}[H]
\begin{center}
\includegraphics[width=10cm]{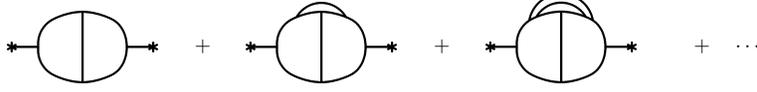}
\end{center}
\caption{Contributions to the shear viscosity from collinear singularities.}
\label{figColl}
\end{figure}

In gauge theories, one also has collinear singularities. 
Fig. \ref{figColl} shows a set of graphs with collinear singularities.
\par

\begin{figure}[H]
\begin{center}
\includegraphics[width=6cm]{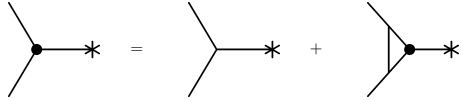}
\end{center}
\caption{An integral equation that resums singularities.}
\label{figBS}
\end{figure}

In general, pinch and collinear singularities are regulated by using hard thermal loop (HTL) resummed propagators. This procedure introduces extra factors of the coupling in the denominators which change the power counting. As a consequence, the infinite series of graphs depicted in Figs. \ref{figPinch} and \ref{figColl} are all of the same order and need to be resummed. The resummation is done by solving a set of coupled integral equations that have  the general form shown in Fig. \ref{figBS}. When the leg on the right hand side has a star,  the integral equation resums pinch singularities. The same equation without the star resums collinear singularities. The basic goal of this paper is to show that these integral equations are produced naturally by the 3PI formalism. We note that since the ghost HTL self energy is zero, the ghost diagram cannot be regulated in this way. However, ghosts are not physical particles and are only needed to cancel unphysical gluon polarizations. We will show explicitly how this works in section \ref{sectionME}.
\par

In Feynman gauge the HTL quark, ghost and gluon propagators can be written:
\bea
&&-i S(P) = \frac{1}{\Psl-\Sigma}\,,~~-i G_{ab}(P) = \frac{\delta_{ab}}{P^2}\,,\\
&& D_{ab}^{\mu\nu}(P) = \delta_{ab}D^{\mu\nu}(P)\,,~~-i D^{\mu\nu}(P)=-P_T^{\mu\nu}\frac{1}{P^2-\Pi_T}-P_L^{\mu\nu}\frac{1}{P^2-\Pi_L}-P_G^{\mu\nu}\frac{1}{P^2}\,,\nonumber\\[2mm]
&&P_T^{\mu\nu} = g^{\mu\nu}-U^\mu U^\nu+(P^\mu-p_0 U^\mu)(P^\nu-p_0 U^\nu)/p^2\,;~~U=(1,0,0,0)\,,\nonumber\\[2mm]
&&P_L^{\mu\nu} = -P_T^{\mu\nu}+g^{\mu\nu}-P^\mu P^\nu/P^2\,,~~
P_G^{\mu\nu} = P^\mu P^\nu/P^2\,.\nonumber
\eea
Dominant contributions to the shear viscosity come from hard excitations on the pinching lines, and the residue of the longtitudinal part of the gluon propagator is exponentially suppressed at high temperatures. Therefore, we need only the transverse part of the pinching gluon propagator which can be written:
\bea
-i D^{ij}_T(P)=\big(\delta^{ij}-\frac{p^i p^j}{p^2}\big)\frac{1}{P^2-\Pi_T}\,.
\eea

Now we consider the vertices defined in (\ref{vertTrans}). We factor the colour structure by defining:
\bea
(\Omega_0)_{ab}^{ij\lambda\tau}=\delta_{ab}\Omega_0^{ij\lambda\tau}\,,~~
(\Theta_0)_{ab}^{ij}=\delta_{ab}\Theta_0^{ij}\,,~~
(\Lambda_0)_{cc^\prime}^{ij}=\delta_{cc^\prime}\Lambda_0^{ij}\,.
\eea
In addition, we write the bare vertices that are obtained directly from the Lagrangian as:  
\bea
\label{vertBare}
&&(\Omega_0)^{\lambda\sigma\tau}_{acb}(-P,0,P)= f_{acb}\Omega_0^{\lambda\sigma\tau}(-P,0,P)\,,\\[2mm]
&&(\Theta_0)^{\sigma}_{acb}(P,0,P)= f_{acb}\Theta_0^{\sigma}(P,0,P)\,,\nonumber\\[2mm]
&&(\Lambda_0)_{cc^\prime}(P,0,P)=\delta_{cc^\prime}\Lambda_0(P,0,P)\,.\nonumber
\eea
Note that we have used the same letters in (\ref{vertTrans}) and (\ref{vertBare}). In order to simplify the notation, we do not introduce any additional primes or tildes to distinguish the two types of vertices. The indices associated with each vertex are sufficient to indicate which type of vertex is meant. Using this notation, it is straightforward to show that: 
\bea
&& \Omega^{\lambda\tau;ij}(-P,0,P) ~~\rightarrow ~~ 2g^{\lambda\tau}p^2 \hat I^{ij} ~~   \rightarrow ~~ p^s \Omega^{\lambda s\tau}(-P,0,P) \hat I^{ij}\,,\\[2mm]
&&\Theta_0^{ij}(P,0,P) = -p^2 \hat I^{ij} = p^s \Theta^s_0(P,0,P) \hat I^{ij} \,,\nonumber \\[2mm]
&&\Lambda^{ij}(P,0,P) = \frac{1}{3}\delta^{ij}p^l\gamma^l-\gamma^i p^j\,,\nonumber
\eea
where the arrow indicates that the relation holds only when multiplied by the transverse projectors $P_T^{\lambda \lambda^\prime}P_T^{\tau\tau^\prime}$ on both sides.
We have taken the limit $Q\to 0$, since this produces no difficulties for factors in the numerator of the integrand. 

Using these results we can rewrite the integrand in (\ref{viscoInt}) in terms of the integrand for the gluon self energy. 
We separate contributions to the gluon self energy from gluon, ghost and quark bubbles by writing:
\bea
\Pi[i]^{s s^\prime}_{ab}(Q) = \int dP\,\Pi int[i]^{ss^\prime}_{ab}(P,Q)\,;~~~i\in\{{\rm gluon},~ {\rm ghost},~ {\rm quark}\}\,.
\eea
It is straightforward to show that (\ref{viscoInt}) can be written:
\bea
\label{etaPi}
\bar\eta = &&\frac{\beta}{10}\hat I^2 \delta_{ab}\int dP\,p^s\,p^{s^\prime}\,\\
&&\cdot\;\bigg(\frac{1}{C_A}\big(\Pi int[{\rm gluon}]^{ss^\prime}_{ab}(P,Q)+\Pi int[{\rm ghost}]^{ss^\prime}_{ab}(P,Q)\big)+\frac{N_c}{d_A T_F}\Pi int[{\rm quark}]^{ss^\prime}_{ab}(P,Q)\bigg)\bigg|_{\scriptsize{
\begin{array}{l}
\vec q =0 \\
q_0\to 0
\end{array}
}}\nonumber
\eea

\section{The 3PI formalism}
\label{section3PI}

At this point, we introduce a compactified notation. We use a single numerical subscript to represent all continuous and discrete indices. For example: a gluon field is written $A_1:=A^a_{\mu}(x)$; 
the quark propagator is written $S_{12}:=S_{\alpha\beta}(x_1,x_2)$; the bare 3-gluon vertex is written $\Omega^0_{132}:=(\Omega_0)^{\lambda\sigma\tau}_{acb}(x_1,x_3,x_2)$ etc.
We also use an Einstein convention in which a repeated index implies a sum over discrete variables and an integration over space-time variables. The  free propagators and vertices are defined as:
\bea
\label{free}
&&(S^0_{12})^{-1}=-i \frac{\delta^2S_{cl}}{\delta\psi_2\delta\bar\psi_1}\,,~~(D^0_{12})^{-1}=-i \frac{\delta^2 S_{cl}}{\delta A_2\delta A_1}\,,~~(G^0_{12})^{-1}=-i \frac{\delta^2 S_{cl}}{\delta \eta_2\delta \bar\eta_1}\,,\\
&&\Lambda^0_{132}=i\frac{\delta^3 S_{cl}}{\delta \psi_2\delta A_3\delta \bar\psi_1}=-\frac{\delta (S^{0}_{12})^{-1}}{\delta A_3}\,,~~
\Omega^0_{132}=i\frac{\delta^3 S_{cl}}{\delta A_2\delta A_3\delta A_1}= -\frac{\delta (D^{0}_{12})^{-1}}{\delta A_3}\,,\nonumber\\
&&
\Theta^0_{132}=i\frac{\delta^3 S_{cl}}{\delta \eta_2\delta A_3\delta \bar\eta_1}= -\frac{\delta (G^{0}_{12})^{-1}}{\delta A_3}\,,~~M^0_{1234}=i\frac{\delta^4 S_{cl}}{\delta A_4 \delta A_3\delta A_2\delta A_1} = \frac{\delta \Omega_{132}^0}{\delta A_4}= -\frac{\delta^2 (D^{0}_{12})^{-1}}{\delta A_4\delta A_3}\,.\nonumber
\eea

The 3PI effective action can be written \cite{berges1,bergesReview}:
\begin{eqnarray}
\label{Gamma3PI}
&&\Gamma[\psi, \bar\psi, A,\eta,\bar\eta, S, D,G,V,U,Y]=S_{cl}[\psi, \bar\psi, A,\eta,\bar\eta]\\[2mm]
&&~~~~+
    \frac{i}{2} {\rm Tr} \,{\rm Ln}D^{-1}_{12}+
\frac{i}{2} {\rm Tr}\left[(D^0_{12})^{-1}\left(D_{21}-D^0_{21}\right)\right]-
  i {\rm Tr} \,{\rm Ln} S^{-1}_{12} - 
i{\rm Tr} \left[(S^0_{12})^{-1}(S_{21}-S^0_{21})\right]\nonumber\\[2mm]
&&~~~~-
  i {\rm Tr} \,{\rm Ln} G^{-1}_{12} - 
i{\rm Tr} \left[(G^0_{12})^{-1}(G_{21}-G^0_{21})\right]+\Gamma^0[A,S,D,G,V,U,Y]+\Gamma^{\rm int}[A,S,D,G,V,U,Y] \,.\nonumber
\eea
We use the following notation:  $V$ is the self consistent quark-gluon vertex, $U$ is the self consistent 3-gluon vertex, and $Y$ is the self consistent ghost-gluon vertex. 
These propagators and vertices are to be determined self-consistently from the equations of motion. 
We also define $\Phi = i(\Gamma^0+\Gamma^{\rm int})$.  We show $\Phi$ graphically  in Fig. \ref{figPhi}. We note that shifting the field introduces another 3-point vertex and that, in the figure, the intersection of three gluon lines at a small dot represents $\Omega_{ijk}^0+A_l M_{ijkl}^0$.
\par\begin{figure}[H]
\begin{center}
\includegraphics[width=10cm]{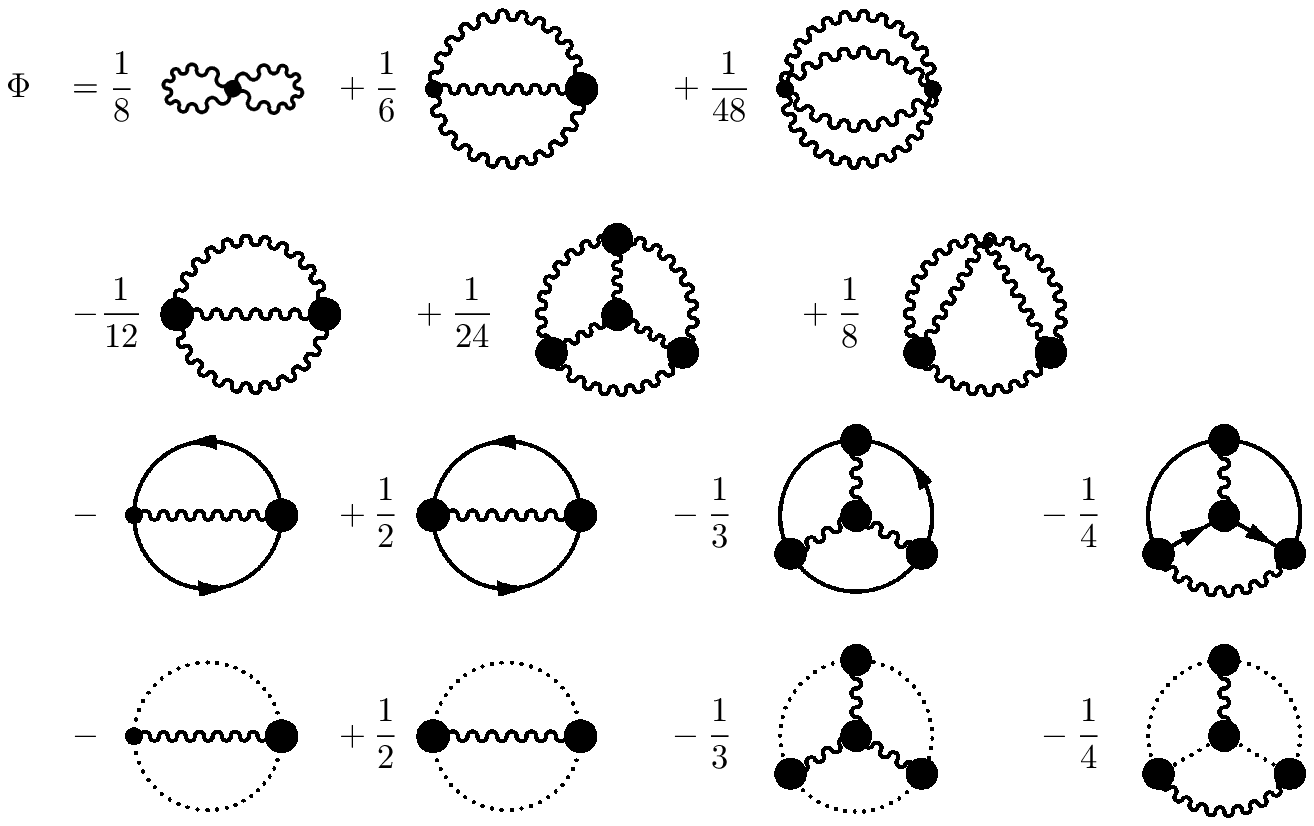}
\end{center}
\caption{The 3-loop 3PI effective action. Wiggly lines represent  gluons, solid lines are quarks and dotted lines are ghosts.}
\label{figPhi}
\end{figure}
The equations of motion are obtained from the stationarity of the action. There are 11 equations which are obtained by functionally differentiating with respect to the 11 functional arguments of the effective action:
\bea
\label{eom1}
&&\frac{\delta \Gamma}{\delta X_i}=0\,;~~~~X_i \in \{\psi,\bar\psi,A,\eta,\bar\eta,S,D,G,U,V,Y\}\,.
\eea
The equations obtained by varying with respect to $\{S,D,G,U,V,Y\}$ can be solved simultaneously for the self consistent solutions which are functions of the field expectation values: $\tilde S[\psi,\bar{\psi},A,\eta,\bar\eta]$, $\tilde D[\psi,\bar{\psi},A,\eta,\bar\eta]$, $\tilde G[\psi,\bar{\psi},A,\eta,\bar\eta]$, $\tilde U[\psi,\bar{\psi},A,\eta,\bar\eta]$, $\tilde V[\psi,\bar{\psi},A,\eta,\bar\eta]$, $\tilde Y[\psi,\bar{\psi},A,\eta,\bar\eta]$.  We write this set of self consistent solutions:
\bea
\label{scSolns}
\tilde X=\{\tilde S,\tilde D,\tilde G,\tilde U,\tilde V,\tilde Y\}\,.
\eea
Substituting these self consistent solutions we 
obtain the resummed action, which depends only on the expectation values of the fields:
\begin{eqnarray}
\label{Gamma3PI-rs}
&& \tilde{\Gamma}[\psi,\bar{\psi},A,\eta,\bar\eta]\\
&&=
\Gamma[\psi,\bar{\psi},A,\tilde{S}[\psi,\bar{\psi},A,\eta,\bar\eta], \tilde{D}[\psi,\bar{\psi},A,\eta,\bar\eta], \tilde{G}[\psi,\bar{\psi},A,\eta,\bar\eta], \tilde{V}[\psi,\bar{\psi},A,\eta,\bar\eta], \tilde{U}[\psi,\bar{\psi},A,\eta,\bar\eta], \tilde{Y}[\psi,\bar{\psi},A,\eta,\bar\eta]]\,.\nonumber
\end{eqnarray}
The equivalence of (\ref{Gamma3PI}) and (\ref{Gamma3PI-rs}) at the exact level was shown in \cite{CJT}. In the future we will write $\Gamma$ and $\tilde \Gamma$ without their arguments.

\section{`External' $n$-point functions}
\label{sectionEXT}

We define some type (2) `mixed' vertex functions
using the same notation as (\ref{free}):
\begin{eqnarray}
\label{vert-defns}
&&\Omega_{132} = -\frac{\delta \tilde{D}^{-1}_{12}}{\delta A_{3}}\,,~~\Theta_{132} = -\frac{\delta \tilde{G}^{-1}_{12}}{\delta A_{3}}\,,~~\Lambda_{132} = -\frac{\delta \tilde{S}^{-1}_{12}}{\delta A_{3}}\,.
\end{eqnarray}
These vertices are shown in Fig. \ref{figExt-V}. `External' gluon legs are distinguished by a star. 
\par\begin{figure}[H]
\begin{center}
\includegraphics[width=8cm]{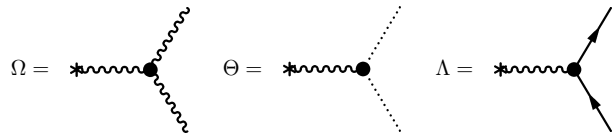}
\end{center}
\caption{Some `mixed' vertices.}
\label{figExt-V}
\end{figure}
\noindent Some additional useful relations can be obtained from the identities:
\begin{eqnarray}
\label{in0}
\tilde{D}^{-1}_{13}\tilde{D}_{32}=\delta_{12}\,,~~\tilde{G}^{-1}_{13}\tilde{G}_{32}=\delta_{12}\,,~~\tilde{S}^{-1}_{13}\tilde{S}_{32}=\delta_{12}\,.
\end{eqnarray}
Differentiating (\ref{in0}) with respect to $A$ and using (\ref{vert-defns}) gives:
\bea
\label{invert}
\frac{\delta\tilde{D}_{12}}{\delta A_{3}} = 
\tilde{D}_{11'}  \Omega_{1'32'}\tilde{D}_{2'2}\,,~~
\frac{\delta\tilde{G}_{12}}{\delta A_{3}} = 
\tilde{G}_{11'}  \Phi_{1'32'}\tilde{G}_{2'2}\,,~~
\frac{\delta\tilde{S}_{12}}{\delta A_{3}}= \tilde S_{11'}\Lambda_{1'3 2'}\tilde{S}_{2'2}\,.
\eea

The `external' gluon propagator is defined as: 
\bea
\label{ext-prop}
i(D^{{\rm ext}}_{12})^{-1}=   
     \frac{\delta^2}{\delta A_2 \delta A_1}
     \tilde{\Gamma}[\psi,\bar{\psi},A,\bar\eta,\eta]\,.
\eea
The `external' self energy is extracted from the `external' propagator using:
\bea
\label{dexternal}
(D^{{\rm ext}}_{12})^{-1}=(D^{0}_{12})^{-1}- \Pi^{\rm ext}_{12}\,.
\eea
We can derive an expression for the `external' self energy as a function of the vertices in (\ref{vert-defns}) by taking derivatives of the modified effective action and using the chain rule. 
We use the notation $X_i$ to indicate one of the set of functional variables $X:=\{S,D,G,V,U,Y\}$ and $\tilde X_i$ to indicate one of the set of self-consistent solutions $\tilde X:=\{\tilde S,\tilde D,\tilde G,\tilde V,\tilde U,\tilde Y\}$. We obtain:
\begin{eqnarray}
\label{Dext-long}
&&i(D^{{\rm ext}}_{12})^{-1} \\
&&= \frac{\delta^2\Gamma}{\delta A_{2} \delta A_{1}}\Big|_{\tilde X}+\sum_i \frac{\delta \Gamma}{\delta X_i}\Big|_{\tilde X}\frac{\delta^2 \tilde X_i}{\delta A_1\delta A_2}
+\Big[\sum_{i}\frac{\delta^2\Gamma}{\delta X_i\delta A_1}\Big|_{\tilde X}\frac{\delta \tilde X_i}{\delta A_2}~+~\{1\leftrightarrow 2\}\Big]
 +\sum_{i}\sum_{j}\frac{\delta^2\Gamma}{\delta X_i\delta X_j}\Big|_{\tilde X}\frac{\delta \tilde X_i}{\delta A_1}\frac{\delta \tilde X_j}{\delta A_2}\,.\nonumber
\end{eqnarray}
The second term in this result is identically zero (see Eqn. (\ref{eom1})). The expression can be further simplified by using the set of equations obtained by differentiating the equations of motion:
\bea
\frac{\delta}{\delta A_2}\;\Big[\frac{\delta \Gamma}{\delta X_i}\Big|_{\tilde X}\Big]=0~~\Rightarrow~~
\frac{\delta^2 \Gamma}{\delta X_i\delta A_2}\Big|_{\tilde X}+\sum_{j}\frac{\delta^2\Gamma}{\delta X_j\delta X_i}\Big|_{\tilde X}\frac{\delta \tilde X_j}{\delta A_2} = 0\,.
\eea
Using this constraint (\ref{Dext-long}) becomes:
\bea
i(D^{{\rm ext}}_{12})^{-1} =\frac{\delta^2\Gamma}{\delta A_{2} \delta A_{1}}\Big|_{\tilde X}+\sum_{i}\frac{\delta^2\Gamma}{\delta X_i\delta A_1}\Big|_{\tilde X}\frac{\delta \tilde X_i}{\delta A_2}\,.
\eea
Expanding the sum and using (\ref{invert}) we have:
\bea
\label{Dext-1}
&& i(D^{{\rm ext}}_{12})^{-1} =\frac{\delta^2\Gamma}{\delta A_{2} \delta A_{1}}\Big|_{\tilde X}\\
&&~~ +
\frac{\delta^2\Gamma}{\delta S_{34}\delta A_1}\Big|_{\tilde X}\cdot(\tilde S_{33'}\Lambda_{3'24'}\tilde S_{4'4})
+\frac{\delta^2\Gamma}{\delta D_{34}\delta A_1}\Big|_{\tilde X}\cdot(\tilde D_{33'}\Omega_{3'24'}\tilde D_{4'4})+\frac{\delta^2\Gamma}{\delta G_{34}\delta A_1}\Big|_{\tilde X}\cdot(\tilde G_{33'}\Theta_{3'24'}\tilde G_{4'4})\nonumber \\[2mm]
&&~~ +\frac{\delta^2\Gamma}{\delta U_{345}\delta A_1}\Big|_{\tilde X}\frac{\delta \tilde U_{345}}{\delta A_2}
+\frac{\delta^2\Gamma}{\delta V_{345}\delta A_1}\Big|_{\tilde X}\frac{\delta \tilde V_{345}}{\delta A_2}
+\frac{\delta^2\Gamma}{\delta Y_{345}\delta A_1}\Big|_{\tilde X}\frac{\delta \tilde Y_{345}}{\delta A_2}\,.\nonumber
\eea
Using (\ref{Gamma3PI}) we obtain:
\bea
\label{Dext-2}
&&-i \frac{\delta^2}{\delta A_2 \delta A_1}\Gamma 
= (D^0_{12})^{-1}-\frac{1}{2}M^0_{1342}\,D_{43}-\frac{\delta^2\Phi}{\delta A_2\delta A_1} \\
&&-i\frac{\delta^2\Gamma}{\delta S_{34}\delta A_1}\Big|_{\tilde X}\cdot(\tilde S_{33'}\Lambda_{3'24'}\tilde S_{4'4}) = \bigg(\Lambda_0 -\frac{\delta^2 \Phi}{\delta S\,\delta A}\bigg)_{413}\cdot(\tilde S\Lambda\tilde S)_{324} \nonumber\\
&&-i \frac{\delta^2\Gamma}{\delta D_{34}\delta A_1}\Big|_{\tilde X}\cdot(\tilde D_{33'}\Omega_{3'24'}\tilde D_{4'4}) =-\frac{1}{2}\bigg(\Omega_0 +2\frac{\delta^2 \Phi}{\delta D\,\delta A}\bigg)_{413}\cdot(\tilde D\Omega\tilde D)_{324} \nonumber\\
&&-i \frac{\delta^2\Gamma}{\delta G_{34}\delta A_1}\Big|_{\tilde X}\cdot(\tilde G_{33'}\Theta_{3'24'}\tilde G_{4'4}) = \bigg(\Theta_0 -\frac{\delta^2 \Phi}{\delta G\,\delta A}\bigg)_{413}\cdot(\tilde G\Theta\tilde G)_{324} \,.\nonumber
\eea
The last three terms in (\ref{Dext-1}) do not contribute because 
the derivatives $\delta \tilde U/\delta A$, $\delta \tilde V/\delta A$, $\delta \tilde Y/\delta A$ correspond to effective 4-point vertices, which are not part of our leading order calculation.
We extract $\Pi_{12}^{\rm ext}$ from (\ref{dexternal}), (\ref{Dext-1}) and (\ref{Dext-2}):
\bea
\label{Dext-3}
\Pi_{12}^{\rm ext} &&= \frac{1}{2}M^0_{1342}\,D_{43}
+\frac{1}{2}\bigg(\Omega_0 +2\frac{\delta^2 \Phi}{\delta D\,\delta A}\bigg)_{413} \cdot(\tilde D\Omega\tilde D)_{324}
-\bigg(\Lambda_0 -\frac{\delta^2 \Phi}{\delta S\,\delta A}\bigg)_{413} \cdot(\tilde S\Lambda\tilde S)_{324} \\
&& -\bigg(\Theta_0 -\frac{\delta^2 \Phi}{\delta G\,\delta A}\bigg)_{413} \cdot(\tilde G\Theta\tilde G)_{324}
+\frac{\delta^2\Phi}{\delta A_2\delta A_1}\,.\nonumber
\eea
The terms in round brackets can be written collectively using the notation: 
\bea
\label{vertprime}
\Omega^\prime_{0}:=\bigg(\Omega_0 +2\frac{\delta^2 \Phi}{\delta D\,\delta A}\bigg)\,,~~
\Lambda^\prime_{0}:=  \bigg(\Lambda_0 -\frac{\delta^2 \Phi}{\delta S\,\delta A}\bigg)\,,~~
\Theta^\prime_{0}:=\bigg(\Theta_0 -\frac{\delta^2 \Phi}{\delta G\,\delta A}\bigg)\,.
\eea
The result for the `external' self energy in (\ref{Dext-3}) is shown in Fig. \ref{piDiag}. The open circles in the figure denote $\Omega_0^\prime$, $\Lambda_0^\prime$ and $\Theta_0^\prime$, and the solid dots are $\Omega$, $\Lambda$ and $\Theta$.
\par\begin{figure}[H]
\begin{center}
\includegraphics[width=17cm]{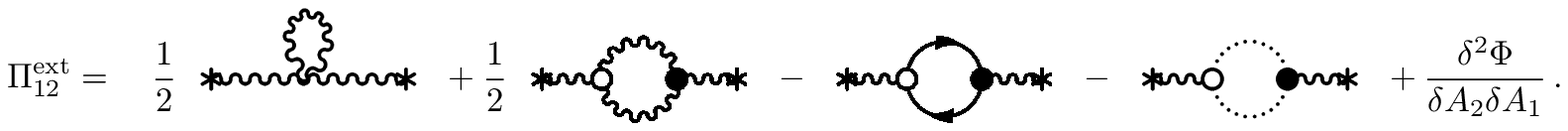}
\end{center}
\caption{The `external' self-energy.}
\label{piDiag}
\end{figure}
\noindent The last term in Fig. \ref{piDiag} can be calculated  using the expression for $\Phi$ shown in Fig. \ref{figPhi}. The result is shown in Fig. \ref{phiDer}. 
\par\begin{figure}[H]
\begin{center}
\includegraphics[width=11cm]{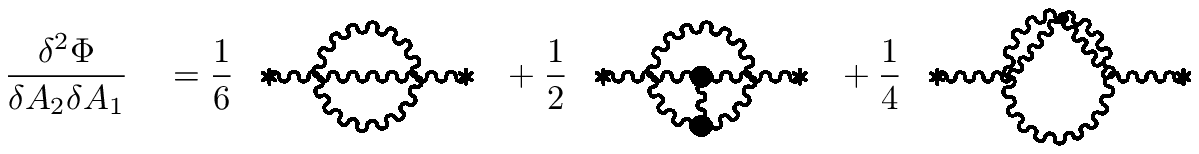}
\end{center}
\caption{A contribution to the conductivity.}
\label{phiDer}
\end{figure}
In order to calculate the viscosity using the 3PI formalism we use Eqn. (\ref{etaPi}), with the integrands for the pieces $\Pi int[{\rm gluon}]^{ss^\prime}_{ab}(K,Q)$, $\Pi int[{\rm quark}]^{ss^\prime}_{ab}(K,Q)$ and $\Pi int[{\rm ghost}]^{ss^\prime}_{ab}$ given by the 2nd, 3rd and 4th terms in (\ref{Dext-3}), which are shown in the 2nd, 3rd and 4th diagrams in Fig. \ref{piDiag}. 
The vertices $\Omega$, $\Theta$ and $\Lambda$, and the self consistent vertices $U$, $V$ and $Y$, satisfy a set of coupled integral equations that resum the pinching and collinear singularities. In the next two sections we derive these integral equations.

In most cases, the equations in this paper are easier to understand when represented diagramatically. From this point on, we will give most results only as diagrams. When equations are used, all indices are suppressed. 

\section{Integral Equations for `Mixed' Vertex Functions}
\label{sectionBS}

In this section we derive the integral equations for the type (2) `mixed' vertex functions $\Omega$, $\Lambda$ and $\Theta$ that appear in the bubble diagrams in $\Pi^{\rm ext}$ (see Fig. \ref{piDiag}).  As explained earlier, these equations are obtained by taking functional derivatives with respect to the field expectation values of the appropriate equations of motion (\ref{eom1}).

The integral equation for the vertex $\Omega$ is obtained from the equation:
\bea
\label{omegaSt}
\frac{\delta}{\delta A}\bigg[\frac{\delta \Gamma}{\delta D}\bigg|_{\tilde X}\bigg]=0\,.
\eea
The subscript $\tilde X$ indicates that all self consistent solutions (\ref{scSolns}) are substituted. 
Using (\ref{Gamma3PI}) and (\ref{vert-defns}) it is straightforward to show that this expression can be written:
\bea
\Omega = \Omega_0^\prime+2\sum_i\frac{\delta \tilde X_i}{\delta A}\,\bigg[\frac{\delta^2\,\Phi}{\delta D\,\delta X_i}\bigg|_{\tilde X}\bigg]\,,
\eea
where the summation indicates contributions from all of the terms in the set $\{S,D,G,V,U,Y\}$. The terms $X_i \in \{U,V,Y\}$ give no contribution, because the derivatives $\delta \tilde X_i/\delta A$ correspond to effective 4-point vertices, which are not part of the leading order calculation. Expanding the sum we have:
\bea
\label{bS1}
\Omega = \Omega^\prime_0
+2\frac{\delta \tilde D}{\delta A}\,\bigg[\frac{\delta^2\,\Phi}{\delta D\delta D}\bigg|_{\tilde X}\bigg] 
+2\frac{\delta \tilde S}{\delta A}\,\bigg[\frac{\delta^2\,\Phi}{\delta D\,\delta S}\bigg|_{\tilde X}\bigg]
+2\frac{\delta \tilde G}{\delta A}\,\bigg[\frac{\delta^2\,\Phi}{\delta D\,\delta G}\bigg|_{\tilde X}\bigg]\,.
\eea
The integral equations for the vertices $\Lambda$ and $\Theta$ are obtained  in exactly the same way. The results are:
\bea
\label{bS2}
&&\Theta = \Theta^\prime_0
-\frac{\delta \tilde D}{\delta A}\,\bigg[\frac{\delta^2\,\Phi}{\delta G\delta D}\bigg|_{\tilde X}\bigg] 
-\frac{\delta \tilde S}{\delta A}\,\bigg[\frac{\delta^2\,\Phi}{\delta G\,\delta S}\bigg|_{\tilde X}\bigg]
-\frac{\delta \tilde G}{\delta A}\,\bigg[\frac{\delta^2\,\Phi}{\delta G\,\delta G}\bigg|_{\tilde X}\bigg]
\,,\\
&&\Lambda = \Lambda^\prime_0
-\frac{\delta \tilde D}{\delta A}\,\bigg[\frac{\delta^2\,\Phi}{\delta S\delta D}\bigg|_{\tilde X}\bigg] 
-\frac{\delta \tilde S}{\delta A}\,\bigg[\frac{\delta^2\,\Phi}{\delta S\,\delta S}\bigg|_{\tilde X}\bigg]
-\frac{\delta \tilde G}{\delta A}\,\bigg[\frac{\delta^2\,\Phi}{\delta S\,\delta G}\bigg|_{\tilde X}\bigg]\,.\nonumber
\eea
To simplify the notation, we define the 4-point functions:
\bea
\label{Mdefn}
M_{SS}:=-\frac{\delta^2\Phi}{\delta S\delta S}\,,~~M_{SD}:=-2\frac{\delta^2\Phi}{\delta D\delta S}\,,~~M_{DD}:=4\frac{\delta^2\Phi}{\delta D\delta D}\,\cdots
\eea
where the dots indicate that the definitions for the 4-point functions involving ghosts are defined like the ones with quark fields. From Fig. \ref{figPhi} it is clear that $M_{SG}=M_{GS}=0$. For clarity, we give one example in Fig. \ref{figM}, with all indices written out explicitly. The legs on each side of the box will join to a pinching pair of propagators. 
\par\begin{figure}[H]
\begin{center}
\includegraphics[width=8cm]{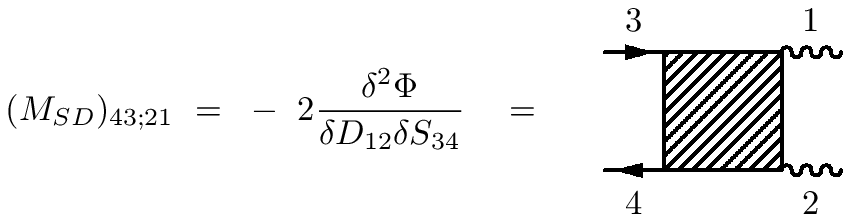}
\end{center}
\caption{A 4-point vertex.}
\label{figM}
\end{figure}

\noindent Using this notation, equations (\ref{bS1}) and (\ref{bS2}) can be represented diagramatically as shown in Fig. \ref{figBSbig}. The shaded boxes in the figure represent the 4-point vertices $M_{XY}$. We use the notation: 
$M_{DD}$ = box with diagonal lines,
$M_{DS}$, $M_{SD}$ = box with hatched lines,
$M_{DG}$, $M_{GD}$ = light grey box,
$M_{SS}$ = dark grey box,
$M_{GG}$ = hatched grey  box.
The open circles denote the vertices $\Omega_0^\prime$, $\Lambda_0^\prime$ and $\Theta_0^\prime$.
\par\begin{figure}[H]
\begin{center}
\includegraphics[width=13cm]{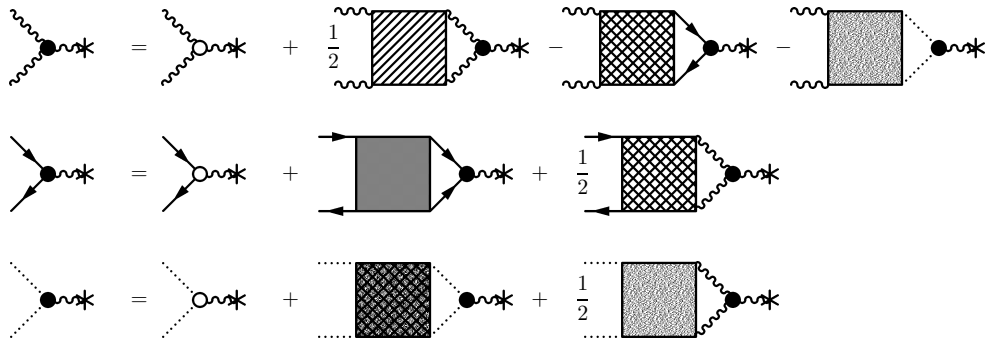}
\end{center}
\caption{Structure of the integral equations for the vertices $\Omega$, $\Lambda$ and $\Theta$.}
\label{figBSbig}
\end{figure}

We can calculate the bare vertices directly from (\ref{vertprime}). Since $\delta^2\Phi/\delta S\,\delta A$ = $\delta^2\Phi/\delta G\,\delta A=0$, we have $\Lambda_0^\prime=\Lambda_0$ and $\Theta_0^\prime=\Theta_0$.
The vertex $\Omega_0^\prime$ is slightly more complicated. After many cancellations, the surviving terms are shown in Fig. \ref{omega0}. We note that $\Omega_0^\prime$ contains $M_0$ and $U$, but not $\Omega$. In Fig. \ref{omega0}, we have combined diagrams that correspond to permutations of external legs. The third and fifth diagrams on the right hand side have a loop insertion on the upper leg. These diagrams should each be drawn as two diagrams, with  symmetry factor 1/2, one with the loop insertion on the upper leg and one with  the loop insertion on the lower leg.
\par\begin{figure}[H]
\begin{center}
\includegraphics[width=11cm]{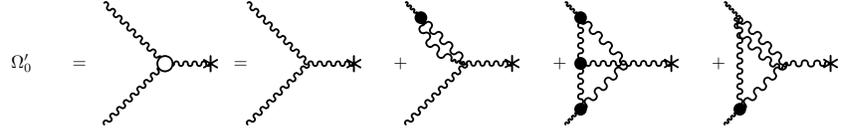}
\end{center}
\caption{The vertex $\Omega_0^\prime$.}
\label{omega0}
\end{figure}

\vspace*{.5cm}

Below we discuss in detail the calculation of $M_{DD}$. In Fig. \ref{figMDD} we show each term in $\Phi$ and the corresponding contribution to $M_{DD}$. As in Fig. \ref{omega0}, we combine diagrams that correspond to permutations of external legs.  The diagram on the right hand side of part (b) in Fig. \ref{figMDD} should be drawn as two diagrams, one with the dotted vertex at the top and one with the dotted vertex at the bottom. Similarly, the diagram in the second part of the right hand side of part (e) should be drawn as two diagrams with the triangular insertion at the top in one diagram and the bottom in the other. In the same way, the first and second diagrams in the right hand side of part (f), and the diagrams in right hand side of (g) and (i) should be drawn as two diagrams. The third diagram in the right hand side of (f) should be drawn as 4 diagrams.
\par\begin{figure}[H]
\begin{center}
\includegraphics[width=12cm]{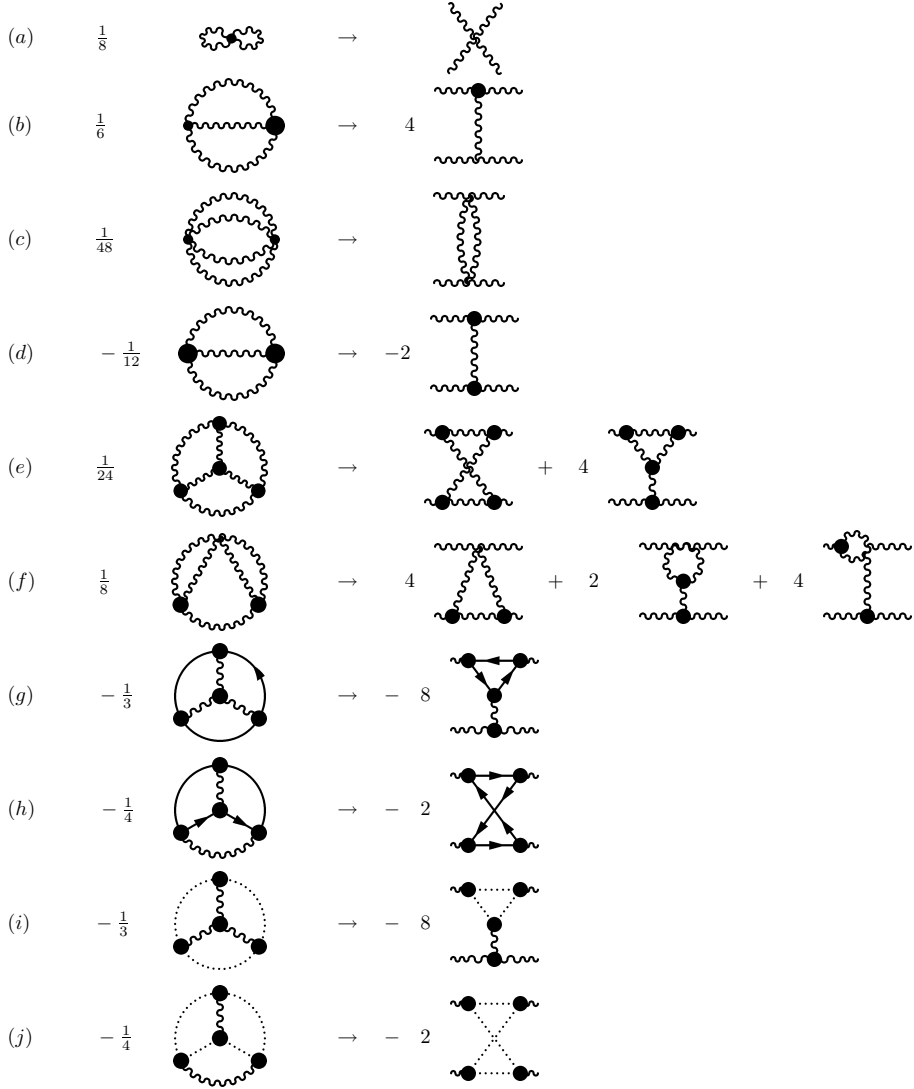}
\end{center}
\caption{Diagrammatic representation of $M_{DD}$.  The crossed lines in parts (e), (h) and (j) pass over/under each other and do not intersect at a 4-point vertex.}
\label{figMDD}
\end{figure}

In Fig. \ref{figMSSD} we give the results for $M_{SD}$ and $M_{SS}$. The corresponding results for $M_{GD}$ and $M_{GG}$ have exactly the same form as $M_{SD}$ and $M_{SS}$ respectively, with the quarks replaced by ghosts.
\par\begin{figure}[H]
\begin{center}
\includegraphics[width=12cm]{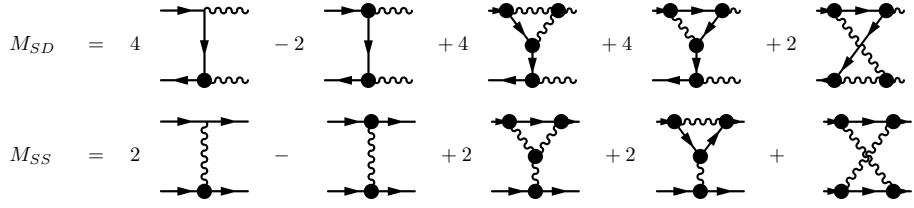}
\end{center}
\caption{Diagrammatic representation of $M_{SD}$ and $M_{SS}$. The crossed lines in the last two diagrams on the right hand side pass over/under each other, and do not intersect at a 4-point vertex. As in Fig. \ref{figMDD}, we have combined diagrams that correspond to permutations of external legs}
\label{figMSSD}
\end{figure}

 The integral equations for the vertices $\Omega$, $\Lambda$ and $\Theta$ are obtained by substituting the equations represented in Figs. \ref{omega0}, \ref{figMDD} and \ref{figMSSD} (and the corresponding equations for ghosts), into Eqns. (\ref{bS1}) and (\ref{bS2}) (shown in Fig. \ref{figBSbig}). \\

\section{Equations of motion for the self consistent vertices}
\label{sectionEOM}

In this section we derive a set of integral equations for the self consistent vertex functions.  We show in detail how each term is obtained for the integral equation for the vertex $U$. For $V$ and $Y$ we give only the final expression. 

The integral equation for the vertex $U$ is obtained from the equation of motion $\delta \Gamma/\delta U=0$. In Fig. \ref{prelimUeqn}  we list the terms in $\Phi$ and the corresponding contributions to the integral equation. 
\par\begin{figure}[H]
\begin{center}
\includegraphics[width=6cm]{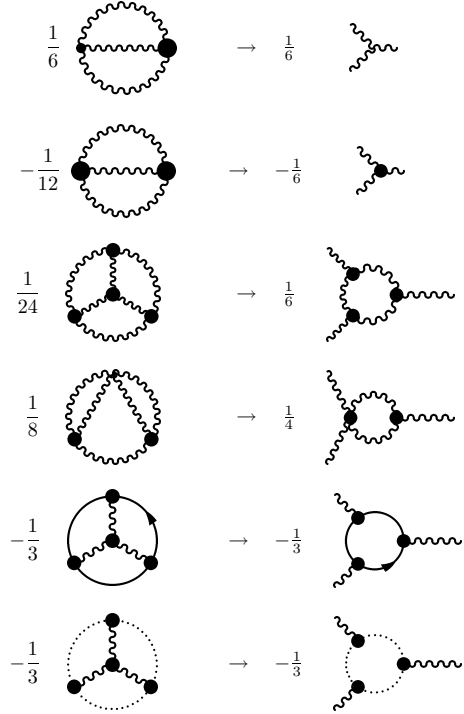}
\end{center}
\caption{Contributions to the integral equation for $U$.}
\label{prelimUeqn}
\end{figure}

\noindent Combining and rearranging, we obtain the integral equation shown in Fig. \ref{Ueqn}.  As before, we have combined diagrams that correspond to permutations of external legs: the third diagram on the right hand side of Fig. \ref{Ueqn} should be drawn as 3 diagrams, each with symmetry factor 1/2.
\par\begin{figure}[H]
\begin{center}
\includegraphics[width=12cm]{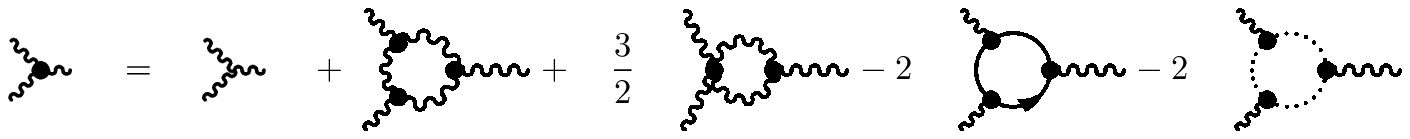}
\end{center}
\caption{The integral equation for $U$.}
\label{Ueqn}
\end{figure}
In exactly the same way, we can obtain integral equations for the vertices $V$ and $Y$. The result for the vertex $V$ is shown in Fig. \ref{VYeqn}. The equation for the vertex $Y$ has the same form with the quarks replaced by ghosts. 
\par\begin{figure}[H]
\begin{center}
\includegraphics[width=8cm]{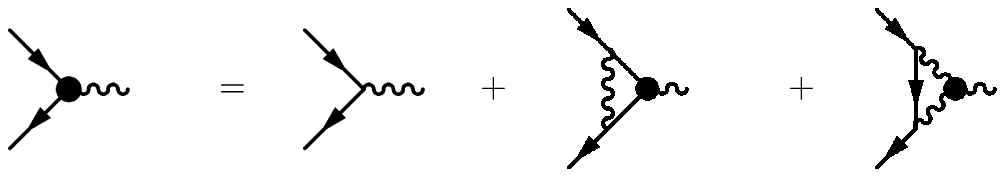}
\end{center}
\caption{The integral equation for $V$.}
\label{VYeqn}
\end{figure}
As a check of our results, we verify that the formalism does not double count the collinear divergences. 
We start by considering the diagrams on the right hand side of Fig. \ref{figMDD}.  We substitute $U_0$ into the diagram in part (b) using Fig. \ref{Ueqn}. The result is to change the sign of diagram (d), and to cancel the second diagram in part (e), the second and third diagrams in part (f), and the diagrams in part (g) and (i). This result is shown in Fig. \ref{figMDDnew}.  The diagrams that have been removed by the substitution are exactly cancelled because they are already contained in the diagrams shown in Fig. \ref{figMDDnew}, together with the integral equations for $U$, $V$ and $Y$, as shown in Figs. \ref{Ueqn} and \ref{VYeqn}.
\par\begin{figure}[H]
\begin{center}
\includegraphics[width=13cm]{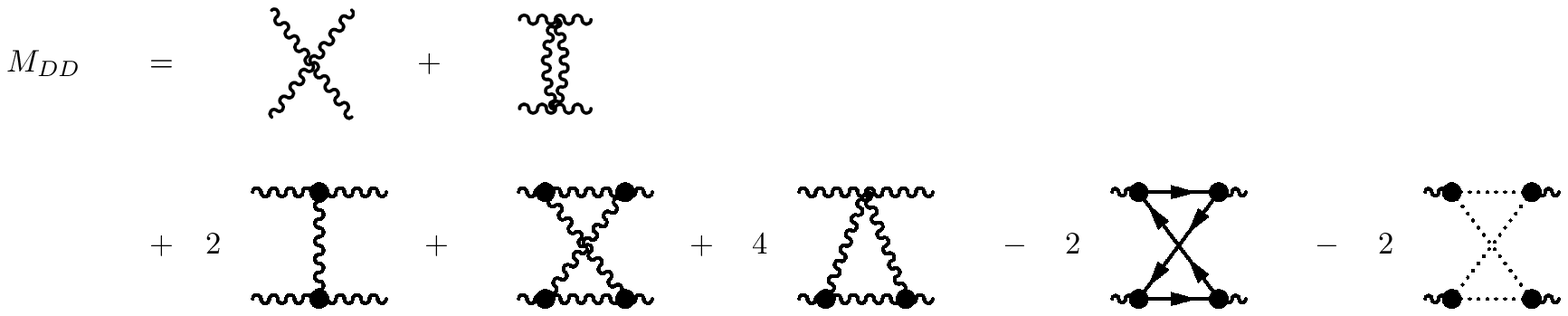}
\end{center}
\caption{A rearrangement of the result for $M_{DD}$.}
\label{figMDDnew}
\end{figure}
Similarly, using Fig. \ref{VYeqn} to remove $V_0$ from Fig. \ref{figMSSD} produces Fig. \ref{figMSSDnew}.
\par\begin{figure}[H]
\begin{center}
\includegraphics[width=5cm]{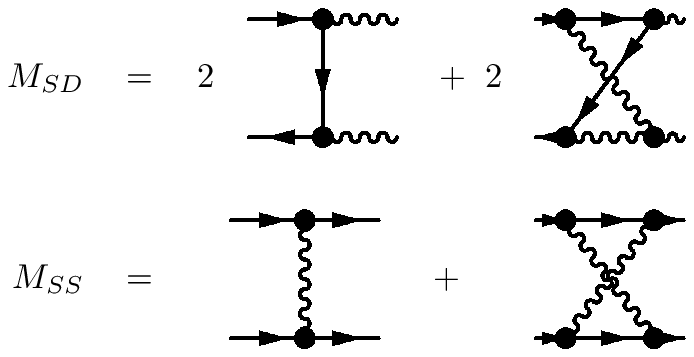}
\end{center}
\caption{A rearrangement of the results for $M_{SD}$ and $M_{SS}$.}
\label{figMSSDnew}
\end{figure}
\noindent Again, we see that the diagrams that have been removed are precisely those that are contained in Fig. \ref{VYeqn}. The cancellation works in exactly the same way for $M_{GD}$ and $M_{GG}$. 

\section{Scattering and Production Processes}
\label{sectionME}

The results of the previous sections can be summarized as follows. 
In order to calculate the leading order QCD shear viscosity using the 3PI formalism we use Eqn. (\ref{etaPi}), with the different pieces of the integrand for the self energy given in (\ref{Dext-3}) (and shown in Fig. \ref{piDiag}). 
The vertices $\Omega$, $\Theta$ and $\Lambda$ and the self consistent vertices $U$, $V$ and $Y$ satisfy a set of coupled integral equations that resum the pinching and collinear singularities. These integral equations are produced naturally by the 3PI formalism and are shown in Figs. \ref{figBSbig}, \ref{Ueqn} and \ref{VYeqn}.

We can demonstrate that the integral equations produced by the 3PI formalism are correct by showing that the kernels of the $\Omega$, $\Theta$ and $\Lambda$ equations have the form of the square of the sum of the amplitudes that correspond to all physical scattering and production processes. In this section we outline the strategy of the calculation. Some details are given in Appendix \ref{sectionOmega}. The basic steps are as follows.\\

{\bf (1)}  We re-expand the equations shown in Fig. \ref{figBSbig}, keeping all terms to 2-loop order. Equivalently, we keep all contributions to the 4-point functions, up to 1-loop order. The expansions of the shaded boxes are shown in Fig. \ref{figMDDnew} and Fig. \ref{figMSSDnew}.  We use Figs. \ref{Ueqn} and \ref{VYeqn} to expand the $U$, $V$ and $Y$ vertex functions, and propagators are expanded by inserting the 1-loop pieces of the self energy (see, for example,  Figs. \ref{piPart1} and \ref{piPart2}). \\

{\bf (2)} We perform the summations over Keldysh indices. The calculation for the $\Lambda$ vertex for QED is done in detail in \cite{MC-EK1}. The method used here is exactly the same. We can write the resulting set of equations in the form:
\bea
\label{BSxform}
&&V_x(P)=V^0_x(P)+\sum_{y\in \{\rm gl,gh,q\}}{\cal S}_{xy}\int dK\, {\cal M}_{xy}(P,K)\Delta^{ret}_y(K) V_y(K) \Delta^{adv}_y(K)\,;~~~x\in \{\rm gl,gh,q\}\,,\\[2mm]
&& V_{\rm gl} = \Omega\,,~~V_{\rm gh} = \Theta\,,~~V_{\rm q} = \Lambda\,,\nonumber\\
&& \Delta_{\rm gl} = D\,,~~\Delta_{\rm gh} = G\,,~~ \Delta_{\rm q} = S\,,~~\nonumber\\
&&M_{\rm gl\,gl}=M_{DD}\,,~~M_{\rm gl\,gh}=M_{DG}\,,~~M_{\rm gl\,q}=M_{DS}\,,~~ {\rm etc}\,,\nonumber
\eea
where `gl' stands for gluon, `gh' stands for ghost, and `q' stands for quark. 
We have supressed all indices, except that we have explicitly written the momentum variables and the integral over the 4-momentum $K$. 
The subscript $q$ refers to a quark of a distinct flavour, and the sum in (\ref{BSxform}) is over each flavour of quark. 
The 3-point vertex is retarded with respect to the middle leg.
The 4-point function is:
\bea
{\cal M}_{xy}(P,K) = M_{xy}(13,P,K)+N_B(K)\Big(M_{xy}(5,P,K)-M_{xy}(9,P,K)\Big)\,,\nonumber
\eea
where the numerical arguments of the 4-point functions indicate Keldysh components. This notation is explained in Appendix A. The factor ${\cal S}_{xy}$ is the symmetry factor of the diagram. 
For example: for $x={\rm gl}$ we have ${\cal S}_{\rm gl\,gl}=1/2; ~~{\cal S}_{\rm gl\,gh}={\cal S}_{\rm gl\,q}=-1$, and Eqn. (\ref{BSxform}) becomes:
\bea
&&\Omega(P)=\Omega^0(P)+\int dK\,\\
&&\big[\frac{1}{2}\,M_{DD}(P,K)D^{ret}(K)\Omega(K)D^{adv}(K)-M_{DG}(P,K)G^{ret}(K)\Theta(K)G^{adv}(K)-M_{DS}(P,K)S^{ret}(K)\Lambda(K)S^{adv}(K)\big]\,,\nonumber
\eea
which is the equation shown in the first line of Fig. \ref{figBSbig}.\\

{\bf (3)}  Since we are only interested in verifying that the correct matrix elements are produced, we use factors for bare propagators in all numerators, but use HTL self energies to regulate pinch singularities in denominators.   We rewrite the pinching pairs of propagators:
\bea
\label{pinch}
&&S^{ret}_{\alpha\beta}S^{adv}_{\alpha^\prime\beta^\prime}=-\Ksl_{\alpha\beta}\Ksl_{\alpha^\prime\beta^\prime}\,\frac{\rho(K)}{2{\rm Im}\hat \Sigma_{ret}(K)} \,,~~\hat\Sigma_{ret}=\frac{1}{2} {\rm Tr}\big(\Ksl \Sigma_{ret}(K)\big)\,,\\
&&D^{ret}_{\mu\nu}D^{adv}_{\lambda\tau}=-g^{\mu\nu}g^{\lambda\tau}\,\frac{\rho(K)}{2{\rm Im}\Pi^T_{ret}(K)}\,,~~G^{ret}G^{adv}=-\frac{\rho(K)}{2{\rm Im}\Pi^T_{ret}(K)} \,,~~\Pi^T_{ret}=P^T_{\mu\nu}\Pi_{ret}^{\mu\nu}\,,\nonumber\\[2mm]
&&-i\,\rho(K)=1/\big(K^2+i \Sign(k_0)\epsilon\big)-1/\big(K^2-i \Sign(k_0)\epsilon\big)\,.\nonumber
\eea
Note that we regulate the pinching singularities from the pair of ghost propagators and the pair of gluon propagators with the transverse part of the gluon polarization tensor. This is justified because of the fact that after all cancellations have been taken into account, only transverse gluons survive. For future use we define:
\bea
&&\Pi_x\,:~~ x\in\{{\rm gl},{\rm gh},{\rm q}\} ~~\rightarrow ~~\Pi_{\rm gl}=\Pi_{\rm gh}=\Pi^T_{ret}\,,~\Pi_{\rm q}=\hat\Sigma_{ret}\,.\nonumber
\eea \\

{\bf (4)} From (\ref{viscoInt}) we need the real part of each $V_x$ and consequently, from (\ref{BSxform}), we need to extract the real part of each 4-point function
$M_{xy}(P,K)$. 
Our method is related to the Cutkosky rules at finite temperature and is described in \cite{MCcutting}. 
Terms with an even number of on shell or `cut' propagators are real. There are no terms with zero cut lines. It is easy to show that terms with four cut lines do not contribute, because it is kinematically forbidden for three on-shell lines to meet at a vertex:
\bea
\label{kine}
\rho\big(\pm(P_1\pm P_2)\big)\rho(P_1)\rho(P_2) = 0\,.
\eea
The conclusion is that all terms must contain two cut lines. However, some terms containing two cut lines are identically zero.  Since the leg momenta $P$ and $K$ are on-shell (because they connect to pinching pairs of propagators) we can use (\ref{kine}) to obtain: 
\bea
\label{PrinL}
\begin{array}{lll}
\rho(K-L)\Delta_x^{ret}(L) ~&\to ~\rho(K-L){\rm Prin}_x(L)\,,~~~~\rho(K-L)\Delta_x^{adv}(L) ~&\to ~\rho(K-L){\rm Prin}_x(L)\,,\\
& & \\
~~~~~~\Delta_x^{ret}(P-K)~&\to~{\rm Prin}_x(P-K)\,,~~~~~~~~~~~~\Delta_x^{adv}(P-K)~&\to~{\rm Prin}_x(P-K)\,,
\end{array}
\eea
where we have defined ${\rm Prin}_x(P) = 1/2\big(\Delta_x^{ret}(P)+\Delta_x^{adv}(P)\big)$.
The result is that all non-zero terms contain two cut lines that effectively divide the diagram into the product of two amplitudes. 

We note that for diagrams where all propagators carry different momenta, the procedure described above is perfectly  straightforward. In diagrams where more than one propagator carries a given momentum, one must be careful to show that potentially dangerous terms that contain the square of a delta function do not appear. The disappearance of these unphysical terms is a well known result due to the KMS condition \cite{lands}.\\

{\bf (5)} For each of the vertices $V_x$, we define a new vertex $\hat V_x$ by contracting each pinching line by the corresponding external leg. To simplify the form of later results, we also divide by the HTL width of the line:
\bea
\label{xhat}
\hat V_{gl}^\mu=\hat \Omega^\mu(P)=\big[-g_{\lambda\tau}\big]\,\Omega^{\lambda\mu\tau}\,\frac{1}{2 {\rm Im}\Pi_T}\,,~~\hat V_{gh}^\mu=\hat \Theta^\mu(P)=\Theta^{\mu}\,\frac{1}{2 {\rm Im}\Pi_T}\,,~~\hat V_{q}^\mu=\hat \Lambda^\mu(P)=\rm Tr\big[\Psl\,\Lambda\big]\,\frac{1}{2 {\rm Im}\hat\Sigma}\,.
\eea
We recall that the ghosts are unphysical degrees of freedom whose only role is to cancel the contributions from the unphysical gluon polarizations. From Fig. \ref{piDiag} we see that it is the combination $\hat V_{\rm gl}-2\hat V_{\rm gh}$ that appears in the viscosity. As a consequence, we will look at the vertices:
\bea
\label{glgh}
\hat V_q \,,~~\hat V_{\bar q} \,,~~\hat V_g = \hat V_{\rm gl}-2\hat V_{\rm gh}\,,
\eea
where the vertex $\hat V_{\bar q}$ is obtained from $\hat V_{q}$ by conjugation. 
To distinguish these vertices from those defined in (\ref{xhat}), we use the indices $\{a,b,\cdots\} \in \{q,\,\bar q,\,g\}$, instead of $\{x,y,\cdots\} \in \{{\rm q},\,\bar {\rm q},\,{\rm gl},\,{\rm gh}\}$.
Our goal is to rewrite the integral equations for the vertices $\hat V_x$ (in Eqn. (\ref{BSxform})) in terms of the vertices $\hat V_a$, and to show that these equations have the correct form, with the kernels given by the square of the sum of the amplitudes that correspond to the relevant 2 $\to$ 2 scattering and production processes. \\

{\bf (6)} In order to obtain the traditional form of the matrix elements, we must label the momenta in a specific way. Each contribution to the 4-point functions $M_{xy}$ has the form of a cut 1-loop amplitude. Each amplitude depends on the two external momenta $P$ and $K$, and one loop momentum variable that is integrated over. We can introduce a second momentum integration by adding a 4-dimensional delta function. We relabel these four momenta by the four variables $\{P,\;P_2,\;L_1,\;L_2\}$ which are defined so that they correspond to the two external momenta, and the two momenta carried by the cut lines (as discussed in step (4) above, there are always two cut lines). In addition, we choose directions so that $P+P_2=L_1+L_2$, which means that the final expression will contain an overall factor $\int dL_1\int dL_2 \;\delta^4\,(P+P_2-L_1-L_2)$. 
In principle, there are 16 terms which correspond to the $2^4$ possible choices for the signs of the 0-components of the momenta on the four on-shell lines. Since $P$ is an external variable, we make the choice $p^0>0$, which leaves eight terms. Only three of these terms correspond to kinematically allowed 2 $\to$ 2 scattering and production processes. For each diagram we write one of these three terms by choosing $\Sign (p^0)=\Sign (p_2^0)=\Sign (l_1^0)=\Sign (l_2^0)$. The terms corresponding to the other two choices can be obtained by making the changes of variables: $P_2\leftrightarrow -L_1$ and $P_2\leftrightarrow -L_2$. 
We define the notation:
\bea
\label{defn-perms}
&&\int dP_2\int dL_1\int dL_2\;\delta^4(P+P_2-L_1-L_2)~\sum_{perms}f(P,P_2;L_1,L_2) \\
&&= \int dP_2\int dL_1\int dL_2\;\delta^4(P+P_2-L_1-L_2)~\bigg(f(P,P_2;L_1,L_2)
+f(P,-L_1;-P_2,L_2)+f(P,-L_2;L_1,-P_2)\bigg)\,.\nonumber
\eea

Using the notation defined in (\ref{pinch}), (\ref{xhat}) and (\ref{defn-perms}) we can rewrite (\ref{BSxform}) as:
\bea
\label{finalForm}
&&\theta(p_0)\,2{\rm Im}\Pi_a\,\hat V_a = \theta(p_0)\,2{\rm Im}\Pi_a\,\hat V^0_a-\theta(p_0) \sum _{perms}\int_{p_2}\cdot\;\bigg[\bigg.\int_{l_1}\int_{l_2}\;\sum_{(i)}\\
&&{\rm{\cal M}}_{(i)}^{ab\rightarrow cd}(P,P_2;L_1,L_2)\,\cdot\,\big({\rm{\cal \widetilde{M}}}_{(i)}^{cd\rightarrow ab}(L_1,L_2;P,P_2)\big)^\dagger\,{\rm N}(m_a l_b i_c j_d)\bigg.\bigg] \,\hat V_b(P_2)\,\cdot\,(2\pi)^4\, \delta^4(P+P_2-L_1-L_2)\,.\nonumber
\eea
The sum over $(i)$ in (\ref{finalForm}) includes the pairs of amplitudes $M\cdot\widetilde{M}$ produced from the cuts of all of the 4-point functions that appear in the expansion of the kernel of the integral equation. 
The indices $\{m_a,\,l_b,\,i_c,\, j_d\}$ take the values b or f, depending on whether the corresponding line is a boson or fermion.  We define:
\bea
&& {\rm N}(ijlm) = {\rm ab}(l,E_{p_2}){\rm em}(i,E_{l_1}){\rm em}(j,E_{l_2})/{\rm em}(m,E_{p})\,,\\[2mm]
&&{\rm em}(b,E_x) = 1+n_b(E_x)\,,~~{\rm em}(f,E_x) = 1-n_f(E_x)\,,~~{\rm ab}(b,E_x) = n_b(E_x)\,,~~{\rm ab}(f,E_x) = n_f(E_x)\,.\nonumber
\eea

In Fig. \ref{boxtri} we give two examples of the way in which cutting a 4-point function produces the product of two amplitudes. The dashed line indicates the cut propagators. The cut box graph gives the square of the $t$-channel. The cut triangle graph gives the product of the  $t$-channel and the $s$-channel.
\par\begin{figure}[H]
\begin{center}
\includegraphics[width=8cm]{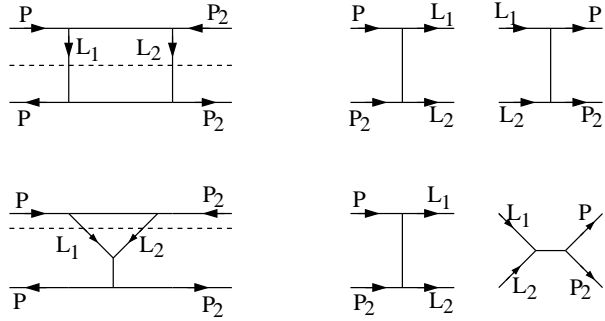}
\end{center}
\caption{Two examples of a cut 4-point funtion written as the product of two amplitudes.}
\label{boxtri}
\end{figure}

The goal is to show that the quantity in square brackets in (\ref{finalForm}) can be rewritten in the form:
\bea
\label{meFF}
\bigg[~~\cdots ~~\bigg]\,\hat V_b(P_2) = \bigg[\frac{1}{\nu_a}\int_{l_1}\int_{l_2}\sum_{\{bcd\}\in\{g\,q\,\bar q\}}\bigg|{\rm{\cal M}}^{ab\rightarrow cd}(P,P_2;L_1,L_2)\bigg|^2\;{\rm N}(m_a l_b i_c j_d)\bigg]\;\hat V_b(P_2)\,.
\eea
The amplitude ${\rm{\cal M}}^{ab\rightarrow cd}$ denotes a scattering amplitude for the process $ab\rightarrow cd$ and the square is summed (not averaged) over the spins and colours of all states. As before, the subscripts $q$ and $\bar q$ denote quarks and anti-quarks of distinct flavours, and the sum is over each flavour of quark. . The factor $\nu_a$ is equal to the number of spin $\times$ colour states for the external excitation, so that dividing by $\nu_a$ produces an average over initial states. The thermal factors give the correct combination of statistical emission and absorption factors. Since we have assumed $\Sign(p_0)>0$,  we obtain the loss term. The choice $\Sign(p_0)<0$ would produce the gain term.  \\

If we take $a=g$ in (\ref{finalForm}) we obtain:
\bea
\label{meResult}
&&\bigg[~~\cdots ~~\bigg]\,\hat V_b(P_2) =\frac{1}{\nu_g}\int_{l_1}\int_{l_2} \\
&&~~\cdot\, \bigg[\big[{\rm {\cal M}}^{gg\to gg}\,{\rm N}({\rm bbbb})+ {\rm {\cal M}}^{gg\to q \bar q}\,{\rm N}({\rm bbf\,f})\big]\hat V_g(P_2)+{\rm {\cal M}}^{gq\to gq}\,{\rm N}({\rm bfbf})\hat V_q(P_2)+{\rm {\cal M}}^{g\bar q\to g\bar q}\,{\rm N}({\rm bfbf})\hat V_{\bar q}(P_2)\bigg]\,,\nonumber\\[4mm]
&& \big|{\bf {\cal M}}^{gg\rightarrow gg}\big|^2=16\,d_A\,C_A^2\, \bigg(3-\frac{st}{u^2}-\frac{su}{t^2}-\frac{tu}{s^2}\bigg)\,,\nonumber\\[2mm]
&& \big|{\bf {\cal M}}^{gg\rightarrow q_1\bar q_1}\big|^2=8 \,d_F\,C_F\, \bigg( C_F\,\bigg(\frac{t}{u}+\frac{u}{t}\bigg)- C_A\,\bigg(\frac{t^2}{s^2}+\frac{u^2}{s^2}\bigg)\bigg)\,,\nonumber\\[2mm]
&&\big|{\bf {\cal M}}^{gq_1\rightarrow g q_1}\big|^2=\big|{\bf {\cal M}}^{g \bar q_1\rightarrow g \bar q_1}\big|^2=8\, d_F\,C_F\,\bigg(C_A\,\bigg(\frac{s^2}{t^2}+\frac{u^2}{t^2}\bigg)-C_F\,\bigg(\frac{s}{u}+\frac{u}{s}\bigg)\bigg)\,.\nonumber
\eea
Taking $a=q$ in (\ref{finalForm}) we obtain:
\bea
\label{meResultq}
&&\bigg[~~\cdots ~~\bigg]\,\hat V_b(P_2) =\frac{1}{\nu_q}\int_{l_1}\int_{l_2} \\
&&~~\cdot\,\bigg[{\rm {\cal M}}^{qq\to qq}\,{\rm N}({\rm f\,f\,f\,f})\hat V_q(P_2) +{\rm {\cal M}}^{qg\to qg}\,{\rm N}({\rm fbfb})\hat V_g(P_2)+\big[{\rm {\cal M}}^{q\bar q\to q \bar q}\,{\rm N}({\rm f\,f\,f\,f})+{\rm {\cal M}}^{q \bar q\to gg}\,{\rm N}({\rm f\,fbb})\big]\hat V_{\bar q}(P_2)\bigg]\,,\nonumber\\[4mm]
&& \big|{\bf {\cal M}}^{q_1 g\rightarrow q_1 g}\big|^2=8\, d_F\,C_F\,\bigg(C_A\,\bigg(\frac{s^2}{t^2}+\frac{u^2}{t^2}\bigg)-C_F\,\bigg(\frac{s}{u}+\frac{u}{s}\bigg)\bigg)\,,\nonumber\\[2mm]
&&\big|{\bf {\cal M}}^{q_1 \bar q_1\rightarrow g g}\big|^2=8 \,d_F\,C_F\, \bigg( C_F\,\bigg(\frac{t}{u}+\frac{u}{t}\bigg)- C_A\,\bigg(\frac{t^2}{s^2}+\frac{u^2}{s^2}\bigg)\bigg)\,,\nonumber\\[2mm]
&& \big|{\bf {\cal M}}^{q_1 q_2\rightarrow q_1 q_2}\big|^2=8\frac{d_F^2\, C_F^2}{d_A}\bigg(\frac{s^2+u^2}{t^2}+\delta_{12}\,\frac{s^2+t^2}{u^2}\bigg)+ 16 \,\delta_{12}\,d_F\, C_F\,\big(C_F-\frac{C_A}{2}\big)\frac{s^2}{t\,u}\,,\nonumber\\[2mm]
&& \big|{\bf {\cal M}}^{q_1 \bar q_2\rightarrow  q_3 \bar q_4}\big|^2=8\frac{d_F^2\,C_F^2}{d_A}\bigg(\delta_{13}\,\delta_{24}\,\frac{s^2+u^2}{t^2}+\delta_{12}\,\delta_{34}\,\frac{t^2+u^2}{s^2}\bigg)+16 \,\delta_{12}\,\delta_{23}\,\delta_{34}\,d_F\, C_F\,\big(C_F-\frac{C_A}{2}\big)\frac{u^2}{s\,t}\,.\nonumber
\eea
The numerical subscripts in Eqns. (\ref{meResult}) and (\ref{meResultq}) refer to quark flavours. 
Our results agree with the $SU(3)$ results of \cite{Combridge}, and with the results of \cite{AMY2}. 
Some details of the calculation of Eqn. (\ref{meResult}) are given in Appendix \ref{sectionOmega}.

\section{Conclusions}
\label{sectionCONC}

In this paper we have presented the first calculation of the complete leading order QCD shear viscosity using quantum field theory methods. 
We have demonstrated that the calculation can be organized naturally using  the 3PI effective action.
The expression produced by the Kubo formula (Eqn. (\ref{etaPi})) contains vertices which satisfy a set of coupled integral equations that resum the pinching and collinear singularities. These integral equations are produced naturally by the 3PI formalism, without the need for any kind of power counting arguments, and are shown in Figs. \ref{figBSbig}, \ref{Ueqn} and \ref{VYeqn}. We have verified that the integral equations produced by the 3PI formalism are correct, by showing that the kernels of the $\Omega$, $\Theta$ and $\Lambda$ equations have the form of the square of the sum of the amplitudes that correspond to all physical scattering and production processes. In principle, the method developed in this paper should be generalizable to the calculation of transport coefficients at higher orders. Work in this direction is in progress. 
Our calculation provides a connection between $n$PI effective theories and kinetic theories, and supports the use of $n$PI effective theories as a method to study the equilibration of quantum fields. \\

\appendix 

\section{Keldysh Representation}
\label{appendixA}

We use the closed time path formulation of real time statistical
field theory \cite{Sch,Keld} which consists of a contour with two
branches: one runs from minus infinity to infinity along the real
axis, the other runs back from infinity to minus infinity just below
the real axis (for reviews see, for example, \cite{gelis,MC-TF}). The closed time path contour results in a doubling of degrees of
freedom.  Physically, these extra contributions come from the
additional processes that are present when the system interacts with a
medium, instead of sitting in a vacuum.  As a result of these extra
degrees of freedom, $n$-point functions have a tensor stucture. 
Statistical field theory
can be formulated in different bases, which produce different
representations of these tensors. We will work in the Keldysh basis. In the discussion below, we use $b_i=1$ or 2 to denote indices in the 1-2 basis and $c_i$ to denote indices in the Keldysh basis, with $c_i = 1 := r$
and $c_i = 2:= a$.
The rotation from the 1-2 representation to the Keldysh representation 
is accomplished by using the transformation matrix:
\begin{equation}
\label{firstU}
U_{Keldysh\leftarrow (1-2) }=\frac{1}{\sqrt{2}}\left(
\begin{array}{lr}
1 & 1 \\
1 & -1
\end{array}
\right).
\end{equation}
The vertices in the Keldysh representation are given by:
\bea
\label{Keldyn}
\Gamma^{c_1\cdots c_n}=2^{\frac{n}{2}-1}\, U^{c_1}\!_{b_1} 
\cdots U^{c_n}\!_{b_n} \Gamma^{b_1\cdots b_n}\,.
\eea
 In order to simplify the notation for the vertices, we replace each combination of the indices $\{r,a\}$ by a single numerical index. In momentum space we write:
\bea
\Gamma^{c_1 c_2\cdots c_n}(p_1,p_2, \cdots p_n) = \Gamma(i,p_1,p_2, \cdots p_n)\,.
\eea 
We assign 
the choices of the variables $c_1 c_2 \cdots c_n$ to the variable $i$ using the vector:
\bea 
\label{ralist}
V_n = 
\Big(
\begin{array}{c}
r_n \\
a_n
\end{array}
\Big) \cdots \otimes
\Big(
\begin{array}{c}
r_2 \\
a_2
\end{array}
\Big)\otimes
\Big(
\begin{array}{c}
r_1 \\
a_1
\end{array}
\Big)\,,
\eea
where the symbol $\otimes$ indicates an outer product. For each $n$,
the $i$th component of the vector corresponds to a list of variables
that is assigned the number $i$.  To simplify the notation we drop the
subscripts and write a list like $r_1 r_2 a_3$ as $rra$. For clarity,
the results are listed below. \\

\xx 3-point functions: $rrr\ra 1$, $arr\ra 2$, $rar\ra 3$, $aar\ra 4$, $rra\ra 5$, $ara\ra 6$, $raa\ra 7$, $aaa\ra
8$,\\

\xx 4-point functions: $rrrr \ra 1$, $arrr \ra 2$,  $ rarr \ra 3$,  $ aarr \ra 4$,  $ rrar \ra 5$,  $ arar \ra 6$,  $ raar \ra 7$,  $aaar \ra 8$,  $ rrra \ra 9$,  $ arra \ra 10$,  $ rara \ra 11$,  $ aara \ra 12$,  $ rraa \ra 13$,  $araa \ra 
14$,  $ raaa \ra 15$, $ aaaa \ra 16$.\\

Summations over Keldysh indices can be done by hand, but the process is extremely tedious. Instead, we use a Mathematica program. This program is described in detail in  \cite{MC-TF} and is available at www.brandonu.ca/physics/fugleberg/Research/Dick.html. The program can be used to calculate the integrand corresponding to any diagram (up to five external legs) in the Keldysh, RA or 1-2 basis. The user supplies input in the form of lists of momenta and vertices for each propagator and vertex.\\

\section{Expansion of the `external' 2-point function}
\label{sectionGI}

As a further check on our calculation, we can show explicitly that to 2-loop order the `external' 2-point function in Eqn. (\ref{dexternal}) (shown in Fig. \ref{piDiag}) contains all terms that one would obtain from a straightforward Wick expansion, with the correct symmetry factors \cite{MC-CP}. 
%
We start with the second term in Fig. \ref{piDiag}. The bare vertex on the left hand side is $\Omega_0^\prime$ as shown in Fig. \ref{omega0}. The dotted vertex on the right hand side is given by the first integral equation in Fig. \ref{figBSbig}, and Figs. \ref{figMDD} and \ref{figMSSD}. 
Term by term we obtain:
\par\begin{figure}[H]
\begin{center}
\includegraphics[width=12cm]{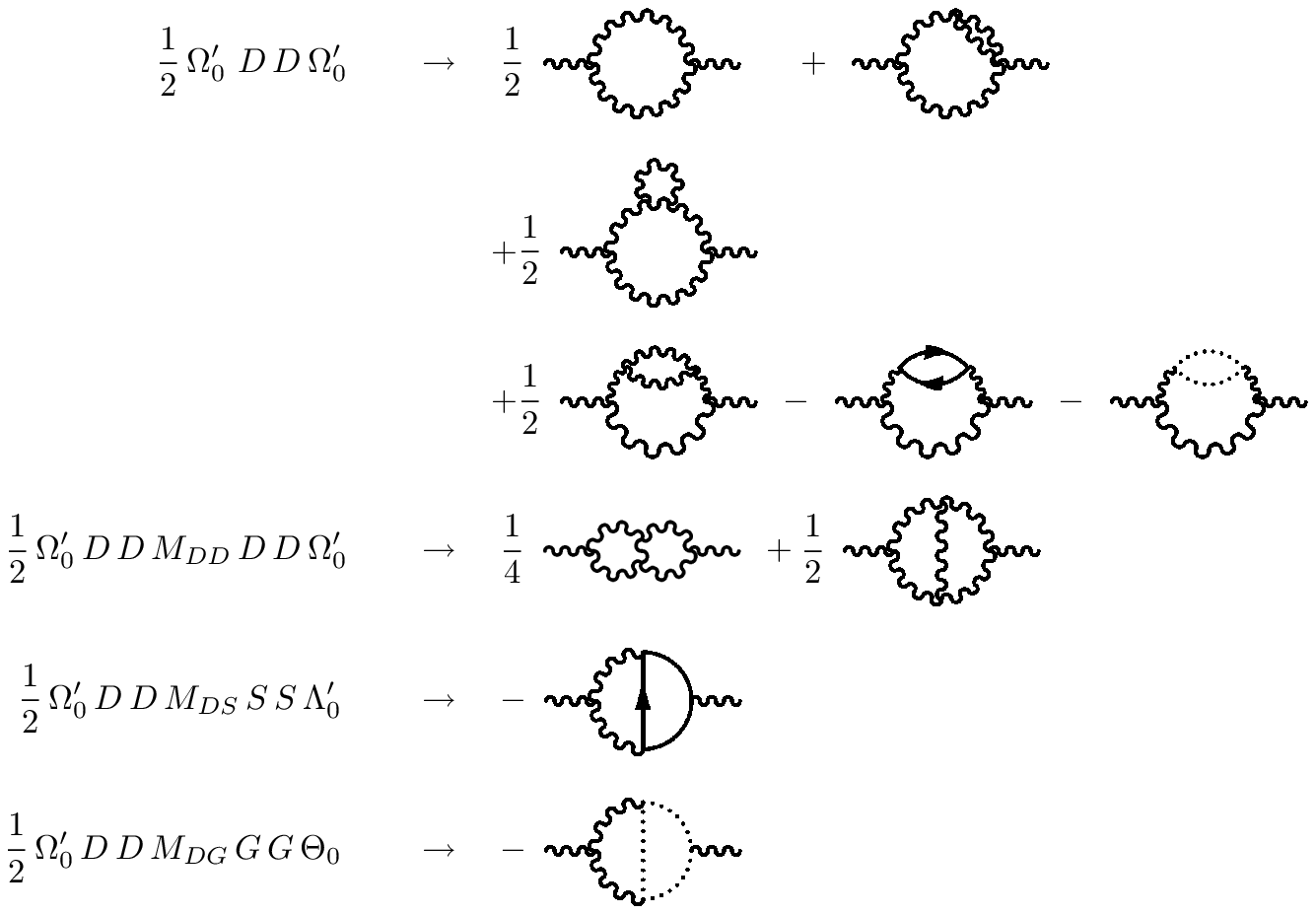}
\end{center}
\caption{Some 2-loop contributions to $\Pi^{\rm ext}$.}
\label{piPart1}
\end{figure}
\xx As discussed before, we have simplified the figure by combining diagrams that correspond to permutations of external legs: the last diagram in the first line of Fig. \ref{piPart1} should be drawn as 2 diagrams, each with a symmetry factor of 1/2, with the loop insertion on the right and left hand side. The diagrams in the second and third lines are obtained by expanding self consistent propagators  and inserting the first diagram in Fig. \ref{piDiag}, and the 1-loop diagrams in Figs. \ref{piPart1}, and \ref{piPart2}. Since we are working to 2-loop order, the full vertices $U$, $V$ and $Y$ in all 2-loop diagrams can be immediately replaced with the bare vertices, using Figs. \ref{Ueqn} and \ref{VYeqn}. \\

Now we look at the third term in Fig. \ref{piDiag}. 
The dotted vertex on the right hand side is given by the second integral equation in Fig. \ref{figBSbig}, and Fig. \ref{figMSSD}. 
Term by term we obtain:
\par\begin{figure}[H]
\begin{center}
\includegraphics[width=9cm]{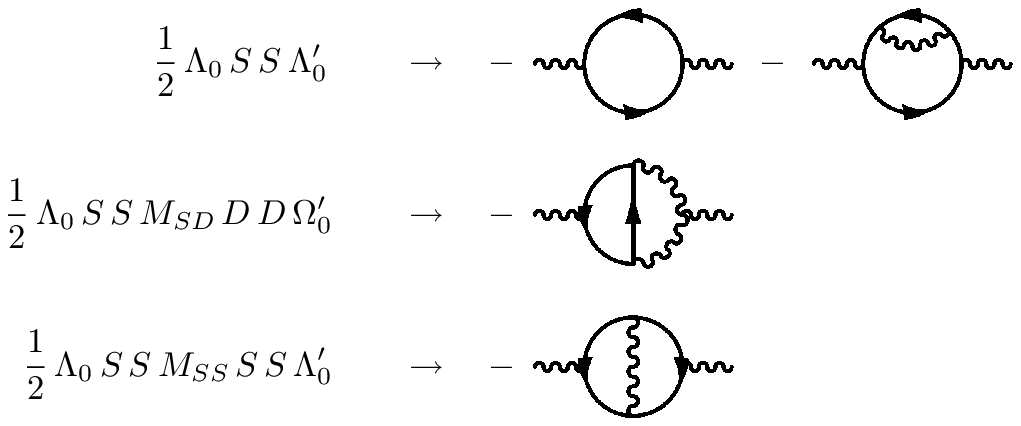}
\end{center}
\caption{Some 2-loop contributions to $\Pi^{\rm ext}$.}
\label{piPart2}
\end{figure}
\noindent The third diagram in Fig. \ref{piDiag} gives the same diagrams as in Fig. \ref{piPart2}, with the quarks replaced by ghosts. \\

To get the full self energy to 2-loop order we combine all contributions. We need the set of diagrams in Figs. \ref{piPart1} and \ref{piPart2} (and the corresponding terms for ghosts).
We also need the tadpole graph in Fig. \ref{piDiag}, and the two loop graph obtained by expanding the self consistent propagator in the tadpole graph and re-inserting the tadpole. This produces the double scoop diagram shown in Fig. \ref{doubleScoop}. Finally, we also need the sunset diagram in Fig. \ref{phiDer}. The symmetry factors are all given explicitly on all diagrams. The resulting set of diagrams is the complete result for the 2-point function, at two loop order. 
\par\begin{figure}[H]
\begin{center}
\includegraphics[width=1.5cm]{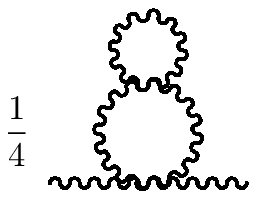}
\end{center}
\caption{The double scoop contribution to $\Pi^{\rm ext}$.}
\label{doubleScoop}
\end{figure}

\section{The $\hat V_g$ equation}
\label{sectionOmega}

In this Appendix we give some details of the calculation of Eqn. (\ref{meResult}). We follow the steps outlined in section \ref{sectionME}. \\

{\bf (1)} We iterate the integral equations and keep all terms up to 2-loop order, or contributions to the 4-point functions up to 1-loop order. These graphs can be divided into types based on their topologies. We call them box, bubble, triangle, cross, loly, tent and fish graphs. 
They are shown in Figs. \ref{boxGraph}, \ref{bubGraph}, \ref{triGraph}, \ref{crossedboxGraph} and  \ref{lolytentfishGraph}, respectively. Each graph carries a numerical factor that is not included in the figure. Graphs that are labeled with the same number, such as $(5)$ and $(5^\prime)$ in Fig. \ref{boxGraph}, give the same result but are drawn separately so that it is easier to see that all contributions are included. 
The numerical factors are listed below, in the same order as the graphs in the corresponding figure. For example: the first line in (\ref{diagFac}) means that the numerical factor for the first box graph is 1, the factor for the second box graph is -2, etc. 
\bea
\label{diagFac}
&&{\rm factor}[{\rm box}]=\{1,-2,-2,-2,-2,-2,-2,-2,-2,4,4,-2\}\,,\\
&&{\rm factor}[{\rm bub}]=\left\{\frac{1}{2},-1,-1,-1,-2,-2,-2,2,2\right\}\,,\nonumber\\
&&{\rm factor}[{\rm tri}]=\{2,-4,-4,-4,-2,-2,-8,-8,-4,-4\}\,,\nonumber\\
&&{\rm factor}[{\rm cross}]=\left\{\frac{1}{2},-1,-1,-2,-2,-2,-2\right\}\,,\nonumber\\
&&{\rm factor}[{\rm loly,\;tent,\;fish}]=\left\{1,2,\frac{1}{2}\right\}\,.\nonumber
\eea
In Figs. \ref{boxGraph}, \ref{bubGraph}, \ref{triGraph} and \ref{lolytentfishGraph}, the dashed lines indicate the two internal propagators that are cut, as discussed in section \ref{sectionME}. For the crossed-box graphs, there are two possible combinations of on shell internal lines which correspond to a horizontal and a vertical cut. In order to simplify Fig. \ref{crossedboxGraph}, we do not draw the dashed lines that correspond to these cuts. 
\par\begin{figure}[H]
\begin{center}
\includegraphics[width=14cm]{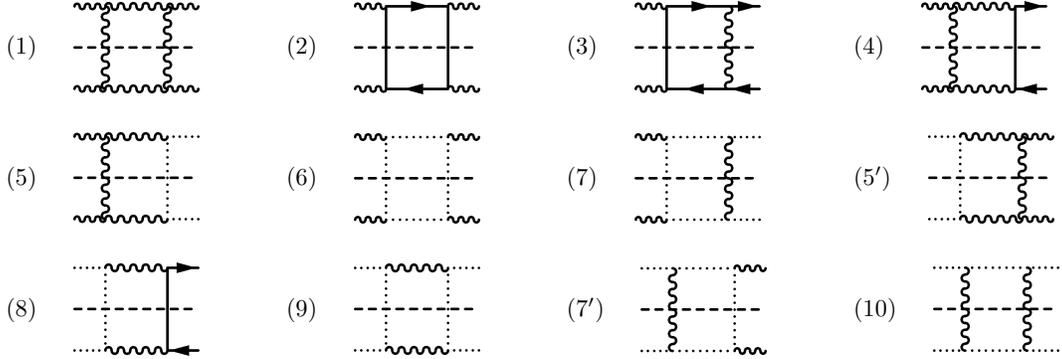}
\end{center}
\caption{Box graphs that contribute to the integral equation for $\hat V_g$.}
\label{boxGraph}
\end{figure}
\par\begin{figure}[H]
\begin{center}
\includegraphics[width=14cm]{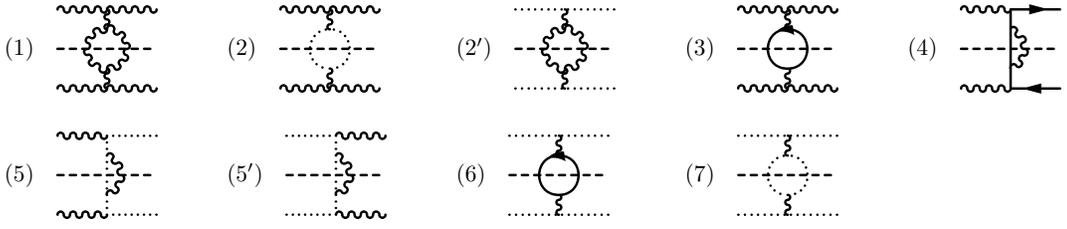}
\end{center}
\caption{Bubble graphs that contribute to the integral equation for $\hat V_g$.}
\label{bubGraph}
\end{figure}
\par\begin{figure}[H]
\begin{center}
\includegraphics[width=14cm]{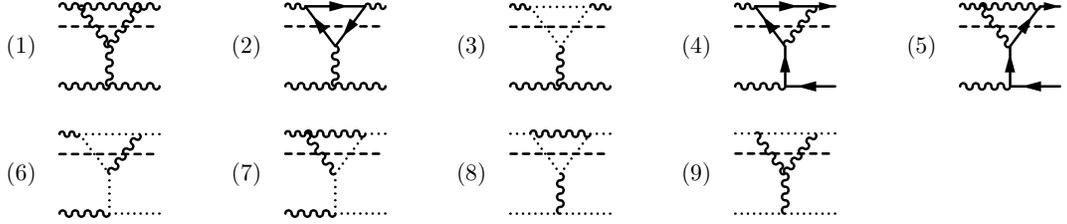}
\end{center}
\caption{Triangle graphs that contribute to the integral equation for $\hat V_g$.}
\label{triGraph}
\end{figure}
\par\begin{figure}[H]
\begin{center}
\includegraphics[width=14cm]{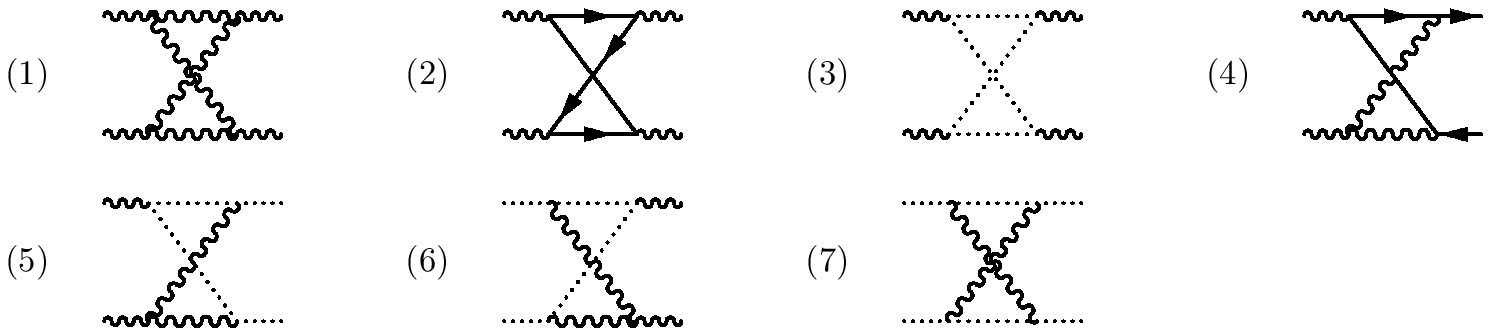}
\end{center}
\caption{Crossed-box graphs that contribute to the integral equation for $\hat V_g$.}
\label{crossedboxGraph}
\end{figure}
\par\begin{figure}[H]
\begin{center}
\includegraphics[width=7cm]{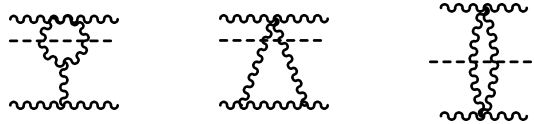}
\end{center}
\caption{Loly, tent and fish graphs that contribute to the integral equation for $\hat V_g$.}
\label{lolytentfishGraph}
\end{figure}

{\bf (2)} We calculate the Keldysh structure for each diagram. In order to do this, we look at the corresponding diagram where all lines are scalars, in the sense that they have no Dirac or Lorentz structure, but carry the appropriate boson or fermion thermal distribution functions. The summations over Keldysh indices are done using a Mathematica program. The program is described in  \cite{MC-TF} and is available at www.brandonu.ca/physics/fugleberg/Research/Dick.html. The program can be used to calculate the integrand corresponding to any diagram (up to five external legs) in the Keldysh, RA or 1-2 basis. The user supplies input in the form of lists of momenta and vertices for each propagator and vertex. Several examples of this part of the calculation are worked out in detail in \cite{MC-EK1}. 
We extract the overall phase space factor:
\bea
{\cal F} =\int_{l_1}\int_{l_2}:= \frac{1}{(2\pi)^3\,2 E_{l_1}}\,\frac{1}{(2\pi)^3\,2 E_{l_2}}\,,
\eea
and define the mandelstam variables:
\bea
s=(L_1+L_2)^2\,;~~t=(P-L_1)^2\,;~~u=(P-L_2)^2\,.
\eea
The results are listed below. The notation `den' is a reminder that the contributions from the numerators, produced by correctly including the appropriate Dirac and Lorentz structure, are not yet included.
\bea
\label{denRes}
&&{\rm den[box(1)]}={\cal F}\, {\rm N}({\rm bbbb})\cdot\frac{1}{t^2}\,;~~i\in\{1,5,6,7,9,10\}\,,\\
&&{\rm den[box(2)]}={\cal F}\, {\rm N}({\rm bbff})\cdot\frac{1}{t^2}\,,~~
{\rm den[box(3)]}={\rm den[box(4)]}={\rm den[box(8)]}=-\,{\cal F}\, {\rm N}({\rm bfbf})\cdot\frac{1}{t^2}\,,\nonumber\\[4mm]
&&{\rm den[bub(1)]}={\cal F}\, {\rm N}({\rm bbbb})\cdot\frac{1}{s^2}\,;~~i\in\{1,2,5,7\}\,,\nonumber\\
&&{\rm den[bub(3)]}={\rm den[bub(6)]}={\cal F}\, {\rm N}({\rm bbff})\cdot\frac{1}{s^2}\,,~~{\rm den[bub(4)]}=-\,{\cal F}\, {\rm N}({\rm bfbf})\cdot\frac{1}{s^2}\,,\nonumber\\[4mm]
&&{\rm den[tri(1)]}={\cal F}\, {\rm N}({\rm bbbb})\cdot\frac{1}{s t}\,;~~i\in\{1,3,6,7,8,9\}\,,\nonumber\\
&&{\rm den[tri(2)]}={\cal F}\, {\rm N}({\rm bbff})\cdot\frac{1}{st}\,,~~{\rm den[tri(4)]}=-\,{\cal F}\, {\rm N}({\rm bfbf})\cdot\frac{1}{st}\,,~~{\rm den[tri(5)]}=-\,{\cal F}\, {\rm N}({\rm bfbf})\cdot\frac{1}{st}\,,\nonumber\\[4mm]
&&{\rm den[cross(1)]}={\cal F}\, {\rm N}({\rm bbbb})\cdot\frac{1}{t u }\,;~~i\in\{1,3,5,6,7\}\,,\nonumber\\
&&{\rm den[cross(2)]}={\cal F}\, {\rm N}({\rm bbff})\cdot\frac{1}{tu}\,,~~{\rm den[cross(4)]}=-\,{\cal F}\, {\rm N}({\rm bfbf})\cdot\frac{1}{tu}\,,\nonumber\\[4mm]
&&{\rm den[loly]}={\cal F}\, i\,{\rm N}({\rm bbbb})\cdot\frac{1}{s}\,,~~{\rm den[tent]}={\cal F}\, i\,{\rm N}({\rm bbbb})\cdot\frac{1}{t}\,,~~{\rm den[fish]}=-{\cal F}\, {\rm N}({\rm bbbb})\,.\nonumber
\eea
The results for the numerators are given in (\ref{numRes}) and are listed in the same order as the diagrams in the corresponding figure. For example, the first line in (\ref{numRes}) gives the numerators for the box graphs labeled (1), (2), etc. The crossed-box graphs produce two contributions each, because there is a horizontal and a vertical cut. 
\bea
\label{numRes}
&&{\rm num}[{\rm box}]=\left\{\right.8 \left(\frac{69 t^2}{2}-25 s u\right) C_A^2,\;-8 t u C_F^2 d_F,\;-8 s t C_F^2 d_F,\;4 \left(4 s^2+3 u s+4 u^2\right) C_A C_F d_F,\\
&&~~~~ -2 \left(t^2+10 s
   u\right) C_A^2,\;2 t^2 C_A^2,\;2 t^2 C_A^2,\;-16 s u C_A^3 C_F d_F,\;4 \left(\frac{s^2}{4}+\frac{u^2}{4}\right) C_A^2,\;2 t^2 C_A^2\left.\right\}\,,\nonumber\\
&&  {\rm num}[{\rm bub}]= \left\{\right.8 \left(\frac{69 s^2}{2}-25 t u\right) C_A^2,\;2 \left(-s^2-10 t u\right) C_A^2,\;4 \left(4 t^2+3 u t+4 u^2\right) C_A C_F d_F,\;-8 s t C_F^2 d_F,2 s^2
   C_A^2,\nonumber\\
&&~~~~ -2 t u C_A C_F d_F,\;4 \left(\frac{t^2}{4}+\frac{u^2}{4}\right) C_A^2\left.\right\}\,,\nonumber\\
&&{\rm num}[{\rm tri}]=\left\{8 \left(15 s u-\frac{3 t^2}{4}\right) C_A^2,\;4 u^2 C_A C_F d_F,\;u^2 C_A^2,0,\;4 s^2 C_A C_F d_F,\,0,\;-s^2 C_A^2,\;t^2 C_A^2,\;-u^2 C_A^2\right\}\,,\nonumber\\
&&{\rm num}[{\rm cross-horz}]=\left\{4 \left(30 t u-\frac{3 s^2}{2}\right) C_A^2,\;0,\;0,\;4 u^2 C_A C_F d_F,\;-u^2 C_A^2,\;u^2 C_A^2,\;0\right\} \,,\nonumber\\
&&{\rm num}[{\rm cross-vert}]=\left\{4 \left(30 t u-\frac{3 s^2}{2}\right) C_A^2,0,0,\;4 u^2 C_A C_F d_F,\;-u^2 C_A^2,\;t^2 C_A^2,\;s^2 C_A^2\right\}\,,\nonumber\\
&&{\rm num}[{\rm loly / tent / fish}]=\left\{324,i \, C_A^2\,s,324\, i\, C_A^2\,s,-864\, C_A^2\right\}\,.\nonumber
\eea

\vspace*{.5cm}

The last step is to sum all contributions and show that:
\bea
\sum_{(i)} {\rm factor}[i]{\rm num}[i]{\rm den}[i] = {\cal F} \big({\rm {\cal M}}^{gg\to gg}\,{\rm N}({\rm bbbb}) + {\rm {\cal M}}^{gg\to q \bar q}\,{\rm N}({\rm bbff})+{\rm {\cal M}}^{gq\to gq}\,{\rm N}({\rm bfbf})+{\rm {\cal M}}^{g\bar q\to g\bar q}\,{\rm N}({\rm bfbf})\big)\,.
\eea
The sum over $(i)$ is over all of the graphs in Figs. \ref{boxGraph}, \ref{bubGraph}, \ref{triGraph}, \ref{crossedboxGraph} and \ref{lolytentfishGraph}.  We use the fact that the definitions of the internal momenta $L_1$ and $L_2$ can always be reversed, or equivalently, that each term can be written in a symmetric form by interchanging $t$ and $u$. Note that when we insert the result for $\big|{\bf {\cal M}}^{ab\rightarrow cd}\big|^2$ into Eqn. (\ref{finalForm}), we must introduce an extra factor 1/2 if the final states $c$ and $d$ are not the same, to avoid double counting this contribution. These factors of two are indicated in square brackets in (\ref{meResult2}). 
The results are:
\bea
\label{meResult2}
&& \big|{\bf {\cal M}}^{gg\rightarrow gg}\big|^2=16\,d_A\,C_A^2\, \bigg(3-\frac{st}{u^2}-\frac{su}{t^2}-\frac{tu}{s^2}\bigg)\,,\\[2mm]
&& \big|{\bf {\cal M}}^{gg\rightarrow q\bar q}\big|^2=\big[\frac{1}{2}\big]\,16 \,d_F\,C_F\, \bigg( C_F\,\bigg(\frac{t}{u}+\frac{u}{t}\bigg)- C_A\,\bigg(\frac{t^2}{s^2}+\frac{u^2}{s^2}\bigg)\bigg)\,,\nonumber\\[2mm]
&&\big|{\bf {\cal M}}^{gq\rightarrow g q}\big|^2=\big|{\bf {\cal M}}^{g \bar q\rightarrow g \bar q}\big|^2=\big[\frac{1}{2}\big]\,16\, d_F\,C_F\,\bigg(C_A\,\bigg(\frac{s^2}{t^2}+\frac{u^2}{t^2}\bigg)-C_F\,\bigg(\frac{s}{u}+\frac{u}{s}\bigg)\bigg)\,,\nonumber
\eea
in agreement with (\ref{meResult}).


\newpage


\begin{thebibliography}{999}
%
\bibitem{GA-review} G. Aarts, PoSLAT2007, 001 (2007) - {\it  arXiv:0811.1850}.
\bibitem{AMY1} P. Arnold, G. Moore and L.G. Yaffe, JHEP {\bf 0011}, 001 (2000) - {\it  arXiv:hep-ph/0010177}.
\bibitem{AMY2} P. Arnold, G.D. Moore and L.G. Yaffe,  JHEP 0301, 030 (2003)   - {\it arXiv:hep-ph/0209353}.
\bibitem{AMY3} P. Arnold, G.D. Moore and L.G. Yaffe,  JHEP 0305, 051(2003)   - {\it arXiv:hep-ph/0302165}.
%
\bibitem{jeonPhi4} S. Jeon, Phys. Rev. {\bf D52}, 3591 (1995) - {\it arXiv:hep-ph/9409250}.
\bibitem{jeonYaffe} S. Jeon and L.G. Yaffe, Phys. Rev. {\bf D53}, 5799 (1996) - {\it arXiv:hep-ph/9512263}.
\bibitem{MC-houRK} M.E. Carrington, D. Hou and R. Kobes, Phys. Rev. {\bf D62}, 025010 (2000) - {\it arXiv:hep-ph/9910344}.
\bibitem{enkeHeinz} E. Wang and U. Heniz, Phys. Rev. {\bf D67}, 025022 (2003) - {\it arXiv:hep-ph/02001116}.
%
\bibitem{basa} M.A. Valle Basagoiti, Phys. Rev. {\bf D66}, 045005 (2002) - {\it arXiv:hep-ph/0204334}.
\bibitem{gertWI} G. Aarts and J.M. Martinez-Resco, JHEP {\bf 0211}, 022 (2002) - {\it arXiv:hep-ph/0209048}.
%
\bibitem{rg} D. Boyanovsky, H.J. deVega and S.Y. Wang, Phys. Rev. {\bf D67}, 065022 (2003) - {\it arXiv:hep-ph/0212107}.

\bibitem{jeonSigma} J.-S. Gagnon and S. Jeon,  Phys. Rev. {\bf D75}, 025014 (2007) - {\it arXiv:hep-ph/0610235}.
\bibitem{jeonEta} J.-S. Gagnon and S. Jeon,  Phys. Rev. {\bf D75}, 025014 (2007) - {\it arXiv:hep-ph/0610235}.
%
\bibitem{gertNf} G. Aarts and J.M. Martinez-Resco, JHEP {\bf 0503}, 074 (2005) - {\it arXiv:hep-ph/0503161}.
%
\bibitem{MC-EK1} M.E. Carrington and E. Kovalchuk, Phys. Rev. {\bf D76}, 045019 (2007) - {\it arXiv:0705.0162}.
%
\bibitem{MC-EK2} M.E. Carrington and E. Kovalchuk, Phys. Rev. {\bf D77}, 025015 (2008) - {\it arXiv:0709.0706}.
%
\bibitem{houQCD} Hou Defu  - {\it arXiv:hep-ph/0501284}.
%
\bibitem{berges1} J. Berges, Phys. Rev. {\bf D70}, 105010 (2004) - {\it  arXiv:hep-ph/0401172}.

\bibitem{smit} A. Arrizabalaga and J. Smit, Phys. Rev. {\bf D66}, 065014 (2002) - {\it arXiv:hep-ph/0301093}.
\bibitem{HZ} M.E. Carrington, G. Kunstatter and H. Zaraket, Eur. Phys. J. {\bf C42}, 253 (2005) - {\it arXiv:hep-ph/0309084}.
%
\bibitem{baym} G. Baym and L. Kadanoff, Phys. Rev. {\bf 124}, 287 (1961).
%
\bibitem{vanh} H. van Hees and  J. Knoll, Phys. Rev. {\bf D66}, 025028 (2002)  - {\it arXiv:hep-ph/0203008}.
%
\bibitem{reinosa1} J. Berges, S. Borsanyi, U. Reinosa and J. Serreau, Annals Phys. {\bf 320}, 344 (2005) - {\it arXiv:hep-ph/0503240}.
\bibitem{reinosa2} U. Reinosa and J. Serreau, JHEP {\bf 0607}, 028 (2006) - {\it arXiv:hep-th/0605023}.
%
\bibitem{serreau} U. Reinosa and J. Serreau, JHEP, {\bf 0711}, 097 (2007) - {\it arXiv:0708.0971}.
%
\bibitem{bergesReview} J. Berges, AIP Conf. Proc. {\bf 739}, 3 (2005) - {\it arXiv:hep-ph/0409233}.
%
\bibitem{CJT} J.M. Cornwall, R. Jackiw and E. Tomboulis, Phys. Rev. {\bf D10}, 2428 (1974).
%
\bibitem{MCcutting} M.E. Carrington, Hou Defu and R. Kobes,  
Phys. Rev. {\bf D67}, 025021 (2003) - {\it arXiv:hep-ph/0207115}.
%
\bibitem{lands} N.P. Landsman and Ch. G. van Weert, Phys. Rep. {\bf 145}, 141 (1987).
%
\bibitem{Combridge} B.L. Combridge, J. Kripfgamz and J. Ranft, Phys. Rev. Lett {\bf B70}, 234 (1977).
%
\bibitem{Sch} P.C. Martin and J. Schwinger, {\it Phys. Rev.} {\bf 115}, 1432 (1959).
\bibitem{Keld} L.V. Keldysh, {\it Sov. Phys. JETP} {\bf 20}, 1018 (1965).
%
\bibitem{gelis} F. Gelis, Nucl. Phys. {\bf B508}, 483 (1997) - {\it arXiv:hep-ph/9701410}.
\bibitem{MC-TF} M.E. Carrington, T. Fugleberg, D.S. Irvine and D. Pickering; 
Eur.Phys.J. {\bf C50}, 711 (2007) -  {\it arXiv:hep-ph/0608298}.
%
\bibitem{MC-CP} C.D. Palmer and M.E. Carrington, Can. J. Phys. {\bf 80}, 847 (2002) - {\it arXiv:hep-th/0108088}.
%
\end{thebibliography}
\end{document}